\documentclass[final,1p,times,authoryear]{elsarticle}

\usepackage{amssymb}
\usepackage{amsthm}
\usepackage{amsmath} 
\newtheorem{theorem}{Theorem}

\usepackage{rotating} 

\journal{Journal of Theoretical Biology}

\begin{document}

\begin{frontmatter}



\title{Analysis of Heterogeneous Cardiac Pacemaker Tissue Models and Traveling Wave Dynamics}


\author{Cheng Ly\corref{cor1}\fnref{label1}}
\ead{cly@vcu.edu}

\author{Seth H. Weinberg\corref{cor1}\fnref{label2}}
\ead{shweinberg@vcu.edu}
\ead[url]{www.shweinberglab.com}

\cortext[cor1]{Corresponding authors}


\address[label1]{Department of Statistical Sciences \& Operations Research,  Virginia Commonwealth University}

\address[label2]{Department of Biomedical Engineering, Virginia Commonwealth University}

\begin{abstract}

The sinoatrial-node (SAN) is a complex heterogeneous tissue that
generates a stable rhythm in healthy hearts, yet a general mechanistic explanation for 
when and how this tissue remains stable is lacking.  Although computational and theoretical analyses could 
elucidate these phenomena, such methods have rarely been used in realistic 
(large-dimensional) 
gap-junction coupled heterogeneous pacemaker tissue models.  
In this study, we adapt a recent model of pacemaker cells \citep{severi12}, incorporating biophysical representations of ion channel and intracellular calcium dynamics, to capture 
physiological features of a heterogeneous population of pacemaker cells, in particular ``center" and ``peripheral" cells with 
distinct intrinsic frequencies and action potential morphology.
Large-scale simulations of the SAN tissue, represented by a heterogeneous tissue structure of pacemaker cells, exhibit a rich 
repertoire of behaviors, including complete synchrony, traveling waves of activity originating from 
periphery to center, and transient traveling waves originating from the center.  
We use phase reduction methods that do not require fully simulating the large-scale model to capture these observations.  
Moreover, the phase reduced models accurately predict key properties of the tissue electrical dynamics, including  wave frequencies when synchronization occurs, 
and wave propagation direction in a variety of tissue models. With the reduced phase models, we analyze the relationship between cell distributions 
and coupling strengths and the resulting transient dynamics.  Further, the reduced phase model predicts parameter regimes of irregular electrical dynamics. Thus, we demonstrate that phase reduced oscillator models applied to realistic pacemaker 
tissue is a useful tool for investigating the spatial-temporal dynamics of cardiac pacemaker activity.

\end{abstract}


\end{frontmatter}


\section{Introduction}
\label{sec:intro}

The sinoatrial node ({\bf SAN}) is the region of the heart responsible for 
generating the electrical rhythm.  
Although isolating individual 
cardiac pacemaker cells in the SAN to experimentally measure 
electrophysiological properties is feasible \citep{honjo96,kodama97}, 
a detailed understanding of the dynamics of intact coupled networks of 
cardiac cells (i.e., tissue) is difficult for several reasons.  The SAN structure is highly complex at both the cellular and tissue levels, 
with heterogeneous cell populations and gap junctional coupling. During normal sinus rhythm, spontaneous electrical activity, driven by the pacemaker cells, propagates throughout the SAN tissue, typically in a center-to-periphery spatial pattern, and then activating the surrounding atrial tissue.  While previous experimental studies have investigated SAN function  
\citep{fedorov09,efimov10,fedorov10,fedorov12,glukhov13,li15,csepe16,li17}, the mechanisms of how regular rhythms are generated in this 
complex tissue structure are still not well understood. Computational modeling can play a valuable role 
in understanding the dynamics of healthy 
and diseased cardiac rhythms 
\citep{greer2017revealing,phadumdeo2018heart,arevalo2016computational,oren2010,zhang2000,wilders07,joyner86}, especially 
as models increase in complexity to incorporate new biological details \citep{severi12}.  

There is some uncertainty as to the degree of cellular heterogeneity within 
the cell population comprising the SAN tissue.  Structural studies have 
described a ``central'' SAN region of smaller cells,  with a transition to 
larger ``peripheral'' SAN cells \citep{bouman86}, with distinct action 
potential and calcium handling properties distinguishing these cell types 
\citep{honjo96,kodama97}.  However, other studies have not found distinct 
isolated SAN cell electrical and calcium properties based on cell size 
\citep{michaels87,verheijck98}. In this study, we will investigate 
electrical properties in the SAN tissue, considering these two different 
approaches: SAN tissue comprised of (1) a homogeneous population of either 
all ``center'' cells or all ``peripheral'' cells (described further below); 
or (2) a heterogeneous population that transitions from center to peripheral 
cells.

While realistic large-scale SAN tissue models can yield insightful simulations, these systems are typically too complicated for detailed mechanistic descriptions of phenomena; a mathematical framework could aid this endeavor. To our knowledge, all computational studies of SAN tissue 
\citep{oren2010,severi12,zhang2000,wilders07,inada14} 
rely on 
tissue-scale simulations, without mechanistic mathematical analysis describing the critical properties that regulate cardiac electrical rhythms.  
There are mathematical studies of simplified SAN models, but without direct correspondence to large-scale biophysical models (see Discussion).  

In this study, we 
apply mathematical methods for studying coupled heterogeneous oscillators (pacemaker cells) to analyze a novel minimal model of SAN tissue electrical activity.  Our coupled oscillator model is developed from the periodic dynamics of a recent multi-compartment biophysical pacemaker model from \citet{severi12}, with realistic calcium dynamics and ionic currents. The Severi et al. model was formulated to address limitations of previous pacemaker cell models, specifically reproducing the effects of perturbations to both ionic currents and calcium handling. Following prior work  \citep{zhang2000,kurata2002,oren2010}, cellular heterogeneity attributes 
(i.e., differences between center and peripheral cells) 
in uncoupled cardiac pacemaker cells are represented by varying model cell size and ionic conductances. We then apply the theory of weakly coupled phase oscillators \citep{kopell86,ermentrout92} to formulate a reduced phase model with a 1D of ``chain'' of coupled phase oscillators, minimally representing the fully coupled SAN tissue in a manner that does 
not require simulating the full tissue model. 

We compare this reduced model with a biophysically detailed SAN tissue model with hundreds of cells, coupled  via gap junctions.  Computationally expensive simulations of several coupled SAN tissue models varying from homogeneous to heterogeneous cell populations 
results in a variety of spatio-temporal dynamics, including: persistent 
synchrony, persistent traveling waves that originate in the periphery, 
transient traveling waves the originate in the center, etc.  Our formulation of a 1D chain of phase oscillators captures the persistent wave frequency, synchronization, and wave propagation direction in a variety of large-scale SAN tissue models.  Furthermore, this mathematical framework can predict the existence and stability of 
traveling waves and characterize the duration of transient activity before 
settling to a persistent state. We demonstrate that this phase reduction 
method can qualitatively and, in some cases, quantitatively 
reproduce electrical properties of biophysically realistic and  
heterogeneous coupled SAN tissue.

\section{Methods}
\subsection*{Detailed Biophysical Pacemaker Cell Models} 

The SAN isolated cell and tissue models are based on the recent rabbit SAN model from \citet{severi12}, with the membrane potential for cell $j$ ($v_j$) governed by the following current-balance equation \citep{oren2010,inada14},
\begin{eqnarray}\label{eqn:volt}
	C_j\frac{dv_j}{dt} & = & -I_{Na}-I_f-I_{to}-I_{Ca,L}-I_{Ca,T}-I_{Kr}-I_{Ks}-I_{Na,K}-I_{Na,Ca} - g_{gap}\sum_k (v_j-v_k),  \nonumber \\	
	 		      &    &
\end{eqnarray}
where $C_j$ is membrane capacitance and which incorporates biophysical descriptions of ionic currents (see Table \ref{tab:currdefn} 
for names of currents), a detailed model of calcium handling including sarcoplasmic reticulum (network and junctional), and ionic current pumps and exchangers.  
There are 31 state variables for \textit{each} pacemaker cell; see \ref{sec:pm_detail} for further details.  
The last term represents local coupling via gap junctions, $g_{gap}\displaystyle\sum_k (v_j-v_k)$, between cell $j$ and cell $k$.  
Gap junctional conductance $g_{gap}$ is only non-zero if cells $j$ and $k$ are coupled and further depend on distance from the tissue center in the heterogeneous coupling schemes considered. Note that Eq.~(\ref{eqn:volt}) assumes that cells are isopotential and that cytoplasmic resistance is neglected since it is much smaller than the gap junctional resistance.

For center cells, we use the same parameters as \citet{severi12} to match the reported action potential characteristics described in \citet{zhang2000,kodama97,oren2010}.  For the other types of 
pacemaker cells (i.e., transitioning from center to peripheral cells, see Fig. \ref{fig:saUncoupled}), we vary the following 12 parameters: cell capacitance, cell length, cell radius, and the maximum conductance of the funny currents, L-type and T-type calcium currents, slow and rapid delayed rectifier potassium currents, sodium current, sodium-potassium pump, and sodium-calcium exchanger. Parameter values for center and peripheral cells (Table \ref{tab:hetparm}) are based on values reported in \citet{oren2010}, and scale factors in \citet{zhang2000}.

A model of the SAN tissue is represented by 271 Severi pacemaker cells with coupling via nearest neighbor gap 
junctions (see Fig. \ref{fig:saIO}{\bf C} left).  
The SAN tissue is represented by a 2D hexagonal grid of 
uniformly spaced cells, i.e., mimicking a radially symmetric architecture, while maintaining the same Euclidean distance between all neighboring cells, to more closely agree with the lower dimensional reduced phase model.  
Simulation of the full biophysical representation of the heterogeneous SAN tissue are computationally expensive, 
requiring several days to simulate 20\,s of biological time for a single 
parameter set.  
The required computational resources (on a High Performance Computing cluster 
with 64-bit AMD computer cores and clock speeds ranging between 2.1 to 2.6 
GHz) are large because the MATLAB (The Mathworks, Natick, MA) stiff ODE solver ({\tt ode15s}) was used with very 
small Relative and Absolute Tolerances of $1\times 10^{-6}$; 
simulations of 20\,s of biological time were used to resolve the periodic state with high
accuracy. Note that histology within the SAN is highly irregular and in our model, cell size varies throughout the hexagonal cell arrangement; however we can estimate the SAN tissue area as $\sim$ 1.5 mm$^2$ (corresponding to a hexagon with a diameter of 19 cells or $\sim$ 1.5 mm), which is smaller than the typical rabbit SAN tissue area of $\sim$ 2 mm x 5 mm, or 10 mm$^2$ \citep{ostborn02}.

\begin{table}[!bht]
\centering
\caption{ Definition of currents in Eq~(\ref{eqn:volt}).}
\label{tab:currdefn}
\begin{tabular}{cl}\hline
{\bf Equation} & {\bf Name}  \\ \hline\hline 
$I_{Na}$ & Sodium current \\ \hline
$I_f$ &  Funny current \\ \hline 
$I_{to}$ &  Transient outward potassium current \\ \hline 
$I_{Ca,L}$ & L-type calcium current \\ \hline 
$I_{Ca,T}$ & T-type calcium current  \\ \hline 
$I_{Kr}$ & Rapid delayed rectifier K$^+$ current \\ \hline 
$I_{Ks}$ & Slow delayed rectifier K$^+$ current \\ \hline 
$I_{Na,K}$ &  Sodium-potassium pump \\ \hline 
$I_{Na,Ca}$ & Sodium-calcium exchanger \\ \hline
\end{tabular}
\end{table}

\begin{sidewaystable}[!bht]
\centering
\caption{ Parameter values for the pacemaker cell model.  All parameters changes proportionally 
($\hbox{min}*(1-s)+\hbox{max}*s$, $s=0$ for Center and $s=1$ for Peripheral) 
except the NaK pump and NaCa exchanger conductances, which were manually altered to insure a 
stable limit cycle (see Table \ref{tab:pumpVals} for these and $s$ values)). See equations in \ref{sec:pm_detail} for 
further details. }
\label{tab:hetparm}
\begin{tabular}{l|c|c|c}\hline
{\bf Parameter} & {\bf Range (Center$\rightarrow$Peripheral) }& {\bf Unit} & {\bf Factor:}  \\ 
& & &  {\bf Peripheral/Center}\\ \hline\hline 
Capacitance, $C$ & [32,65]  & pico-Farads (pF) & 2.03 \\ \hline
Cell Length & [70,86] & micrometer ($\mu$m) & 1.23 \\ \hline 
Cell Radius & [4,6] & micrometer ($\mu$m) & 1.5 \\ \hline 
Na$^+$ (funny current) Conductance, $g_{f,Na}$ & [0.03,0.378] & micro-Siemens ($\mu$S)  & 12.6 \\ \hline 
K$^+$ (funny current) Conductance, $g_{f,K}$ & [0.03,0.378]  & micro-Siemens ($\mu$S)  & 12.6 \\ \hline 
Ca$^{+2}$ (L-type) Conductance, $PCaL$ & [0.2,2.28] & nano Amps per nM (nA/nM) & 11.4 \\ \hline 
Ca$^{+2}$ (T-type) Conductance, $PCaT$ & [0.02,0.064] & nano Amps per nM (nA/nM) & 3.2 \\ \hline 
Rapid K$^+$ Rectifier Conductance, $g_{Kr}$ & [2.1637$\times 10^{-3}$,5*2.1637$\times 10^{-3}$] &  micro-Siemens ($\mu$S) & 5 \\ \hline 
Slow K$^+$ Rectifier Conductance, $g_{Ks}$ & [1.6576$\times 10^{-3}$,5*1.6576$\times 10^{-3}$]& micro-Siemens ($\mu$S) & 5 \\ \hline 
Na$^+$ Conductance, $g_{Na}$ & [0.0125,0.25] & micro-Seimens ($\mu$S) & 20 \\ \hline 
NaK Pump Conductance, $g_{NaK}$ & [0.063,0.4095] & nano-Amps (nA) & 6.5 \\ \hline 
NaCa Exchanger Conductance, $g_{NaCa}$ & [4,40] & nano-Amps (nA) & 10 \\ \hline
\end{tabular}
\end{sidewaystable}

\begin{table}[!bht]
\centering
\caption{Parameter values for NaK pump and NaCa exchanger current and $s$ proportionality scalar $\hbox{min}*(1-s)+\hbox{max}*s$.}
\label{tab:pumpVals}
\begin{tabular}{c|c|c|c}\hline
{\bf Cell Type} & {\bf $g_{NaK}$} (nA) & {\bf $g_{NaCa}$} (nA)  & $s\in [0,1]$ \\ \hline\hline 
1 (center) & 0.063 & 4 & 0  \\ \hline
2 & 0.0695 & 6 & 0.25/9 \\ \hline
3 & 0.08 & 6.7 & 0.65/9 \\ \hline
4 & 0.095 & 8.9 & 1.2/9 \\ \hline
5 & 0.125 & 13.5 & 2/9 \\ \hline
6 & 0.164 & 17 & 3/9 \\ \hline
7 & 0.2045 & 21 & 4/9 \\ \hline
8 & 0.2871 & 29 & 6/9 \\ \hline
9 (peripheral) & 0.4095 & 40 & 1 \\ \hline
\end{tabular}
\end{table}

\subsection*{Phase Reduced Model}

To mathematically analyze the persistent (or steady-state traveling wave) 
behavior of the coupled pacemaker cells, we employ phase 
reduction methods, in which the dynamics of high dimensional systems are approximated with a simple scalar periodic variable.  
These methods have been successfully applied to 
many areas of science \citep{winfree67}, including 
chemical oscillations \citep{kuramotoBook}, 
synchronization of fireflies \citep{mirollo90}, 
circadian rhythms \citep{winfree70,zeitzer00}, and 
cellular networks 
\citep{ermentroutBook,prcNeuroBook,lyErm_siads_10}, to name a few.  We briefly describe the general approach here 
(see Chapter 8 of \citet{ermentroutBook}).  Consider a large dimensional model 
$X_j\in\mathbb{R}^n$ coupled to $X_k\in\mathbb{R}^n$ of the form:
\begin{equation}\label{Xj_hiD}
	\frac{dX_j}{dt} = F(X_j) + \varepsilon G(X_j,X_k).
\end{equation}
When there is an asymptotically stable limit cycle $X_0(t)$ with period $T_j$, for the uncoupled case $\varepsilon=0$, 
we map values near the limit 
cycle to the unit circle via a function $\phi(X_j):=\Theta_j\in[0,T_j)$, to derive the following:
\begin{equation}
	\frac{d\Theta_j}{dt} = 1 +\nabla_X \phi(X_j)\cdot \varepsilon G(X_j,X_k).
\end{equation}
Since other cells $X_k$ only directly 
effect the voltage variable, all components of $\nabla_X \phi(X_j)\cdot \varepsilon G(X_j,X_k)$ 
are 0 except the voltage component.  For small $\varepsilon$, we approximate 
$\nabla_X \phi(X_j(\Theta_j))\approx\nabla_X\phi(X_0(\Theta_j))$; the voltage component of this is the 
infinitesimal \textit{phase-resetting curve} ({\bf PRC}), which we denote by $\Delta_j(\Theta_j)$. Next, the method 
of averaging is applied to approximate the second term of Eq (\ref{Xj_hiD}) with the function:
\begin{equation}\label{eqn:Hfunc}
	H_{j,k}(\Theta_k-\Theta_j):=\frac{1}{T_j}\int_0^{T_j} \Delta_j(t)G\left(X_j(t),X_k(t+\Theta_k-\Theta_j)\right)\,dt,
\end{equation}
often called the interaction function (or $H-$function).  Note that 
PRC of cells can be experimentally measured by injecting pulses at various phases and recording the 
times to next spike (see \citet{sano78,jalife80} who measured PRCs in SAN pacemaker cells of mammals); the 
infinitesimal PRC is the limit of small (i.e., zero) duration and amplitude pulses.

This phase reduction can be applied repeatedly for each of the different pacemakers, 
with a rescaling of time so that $\theta_j\in[0,1)$ to arrive at the following phase network 
approximation:
\begin{equation}\label{eqn:genPhaseMod}
	\frac{d\theta_j}{dt} = \omega_j + \sum_k H_{j,k}(\theta_k-\theta_j)
\end{equation}
Since we are interested in the direction of the traveling wave (if it exists) and the large SAN model 
is nearly radially symmetric in cell type, 
the phase model is further simplified to a 
1-dimensional chain 
with ``cut-ends", with one end being the Center and the other being the Peripheral (see Fig. \ref{fig:phase1}{\bf C}).  Furthermore, a 
relatively large number of cell types (9) are used to span the range from a prototypical 
center pacemaker to peripheral pacemaker, enabling us to assume a modest 
amount of heterogeneity 
among neighboring cells (see Fig. \ref{fig:saUncoupled}{\bf B}).  
Since coupling is local via gap junctions, we approximate the interaction function (for $k=j-1$ and $k=j+1$ only):
\begin{equation}\label{eqn:H_apprx}
	H_{j,k}(x) \approx H_j(x) 
\end{equation}
where $H_j$ is the interaction function of the $j^{th}$ cell type with itself, which can be numerically computed with the program XPP \citep{xppBook}.  
Specifically, we first compute the infinitesimal PRC $\Delta_j(t)$ in Eq.\eqref{eqn:Hfunc} via XPP (the mechanics of which 
are described in Chapter 9.5.1 of \citet{xppBook}), after which the H-function is computed with an automated subroutine in XPP (described in Chapter 9.5.2 of 
\citet{xppBook}).  Note that the required XPP file is publicly available at 
http://github.com/chengly70/SanHeteroSeveri.

In order to systematically relate the 1D chain of phase oscillators to the large scale model, 
each of the cell types 
are grouped together, and an average conductance is calculated via:  
$${\bar g_{gap,j}} = \frac{1}{\hbox{\# nonzero }g_{j,k}} \sum_k g_{j,k} .$$  
Here, $g_{j,k}$ denotes the $g_{gap}$ conductance (see Eq.~(\ref{eqn:volt})) between cell $j$ and cell $k$ (Fig. \ref{fig:saIO}{\bf C} left color-codes all 9 cell types by spatial location).  
This value of ${\bar g_{gap,j}}$ multiplies the corresponding pieces of the interaction functions (plotted in Fig. \ref{fig:phase1}{\bf B}):
\begin{equation}\label{eqn:H_apprx2}
	H_{j,k} \approx  {\bar g_{gap,j}}H_j .
\end{equation} 

We can use this phase reduction approach to determine the existence of traveling wave solutions.  These solutions to equation (\ref{eqn:genPhaseMod}) are of the form:
$$ \theta_j(t)= \Omega t + \xi_j .$$

\subsection*{Quantifying Transient Times with Phase Models}

If a stable persistent state exists for a given phase reduction and tissue structure, we are also interested in the time required for the system to reach this persistent state, following a perturbation or from a given initial condition. We can calculate this transient time in the full SAN tissue model via simulation and compare with the reduced phase model framework.  In the reduced phase model, consider a perturbation 
$\vec{\eta}(t)$ to a traveling wave solution 
$\Omega t+\vec{\xi}$
\begin{equation}
	\theta_j(t)=\Omega t + \xi_j + \eta_j(t).
\end{equation}
After substituting the above equation into 
Eq.~(\ref{eqn:genPhaseMod}) and assuming 
perturbations are small $\Big( \eta_j=O(\varepsilon)$ for all $j$, different $\varepsilon$ than in Eq.(\ref{Xj_hiD})\Big), 
the 1$^{st}$ order $\varepsilon$ equation 
provides a linear approximation for the dynamics of $\eta_j(t)$:  
\begin{eqnarray}\label{eqn:odeEta}
    \frac{d\eta_1}{dt} &=& H'_{1,2}(\xi_2-\xi_1)(\eta_2-\eta_1) \nonumber \\
    \frac{d\eta_j}{dt} &=& H'_{j,j-1}(\xi_{j-1}-\xi_j)(\eta_{j-1}-\eta_j)+H'_{j,j+1}(\xi_{j+1}-\xi_j)(\eta_{j+1}-\eta_j); \hbox{  for }j= 2,\dots, 8 \nonumber \\
    \frac{d\eta_{9}}{dt} &=& H'_{9,8}(\xi_{8}-\xi_{9})(\eta_{8}-\eta_{9})
\end{eqnarray}
This equation in matrix-vector form is:
\begin{equation}\label{eqn:etaMatr}
	\frac{d\vec{\eta}}{dt} = A \vec{\eta}
\end{equation}
\begin{equation}\label{eqn:Adefn}
A:=\begin{pmatrix}
    -H'_{1,2} & H'_{1,2} & 0 & \dots & \dots & \dots  & 0 \\
    \vdots & \vdots & \vdots & \vdots  & \vdots & \vdots  \\
    0 & \cdots 0 & H'_{j,j-1} & -H'_{j,j-1}-H'_{j,j+1} & H'_{j,j+1}  & 0\cdots & 0 \\
    \vdots & \vdots & \vdots & \vdots  & \vdots & \vdots  \\
    0 & \dots & \dots &\dots & 0 & H'_{9,8} & -H'_{9,8}
\end{pmatrix}
\end{equation}
Each of the terms in the matrix abbreviated as: $H'_{j,k}:=H'_{j,k}(\xi_k-\xi_j)$.  
When a traveling wave solution is stable, all but one of the eigenvalues of $A$ 
have negative real part; there is a 
a distinct zero eigenvalue $\lambda_0=0$ with eigenvector $\vec{1}$ that corresponds to a constant shift in 
the traveling wave solution\footnote{A constant shift in all variables is a perturbation that does not decay, because the result is the same traveling wave solution.}.  

Let the eigenvectors of $A$ be $\vec{\psi}_j$ with eigenvalues $\lambda_j$: $A\vec{\psi}_j = \lambda_j\vec{\psi}_j$, then since 
$A$ is symmetric, we take $\vec{\psi}_j$ to be an orthonormal set.  With initial condition $\vec{\eta}_0$, 
the vector of perturbations is captured 
using an eigenvector expansion to solve for:
$$\frac{d\vec{\eta}}{dt} = A \vec{\eta} $$
to get:
$$\vec{\eta}(t)=  \sum_{k\neq 0}(\vec{\psi}_k,\vec{\eta_0}) e^{\lambda_k t} \vec{\psi_k}.$$
The solution can be approximated by using only the largest real eigenvalue with an eigenvector expansion 
(excluding the 0 eigenvector/value): 
$$\vec{\eta}(t)=  (\vec{\psi}_1,\vec{\eta_0}) e^{\lambda_1 t} \vec{\psi_1}.$$
The operation $(\vec{v},\vec{w})$ is the usual inner product.  
Fig S2 (S1Text.pdf) shows that this approximation is excellent for the 
tissues and initial condition we consider.  

\section{Results}

\subsection{Large-scale Pacemaker Tissue}

\begin{figure}[!htb]
\begin{center}
\includegraphics[width=\textwidth]{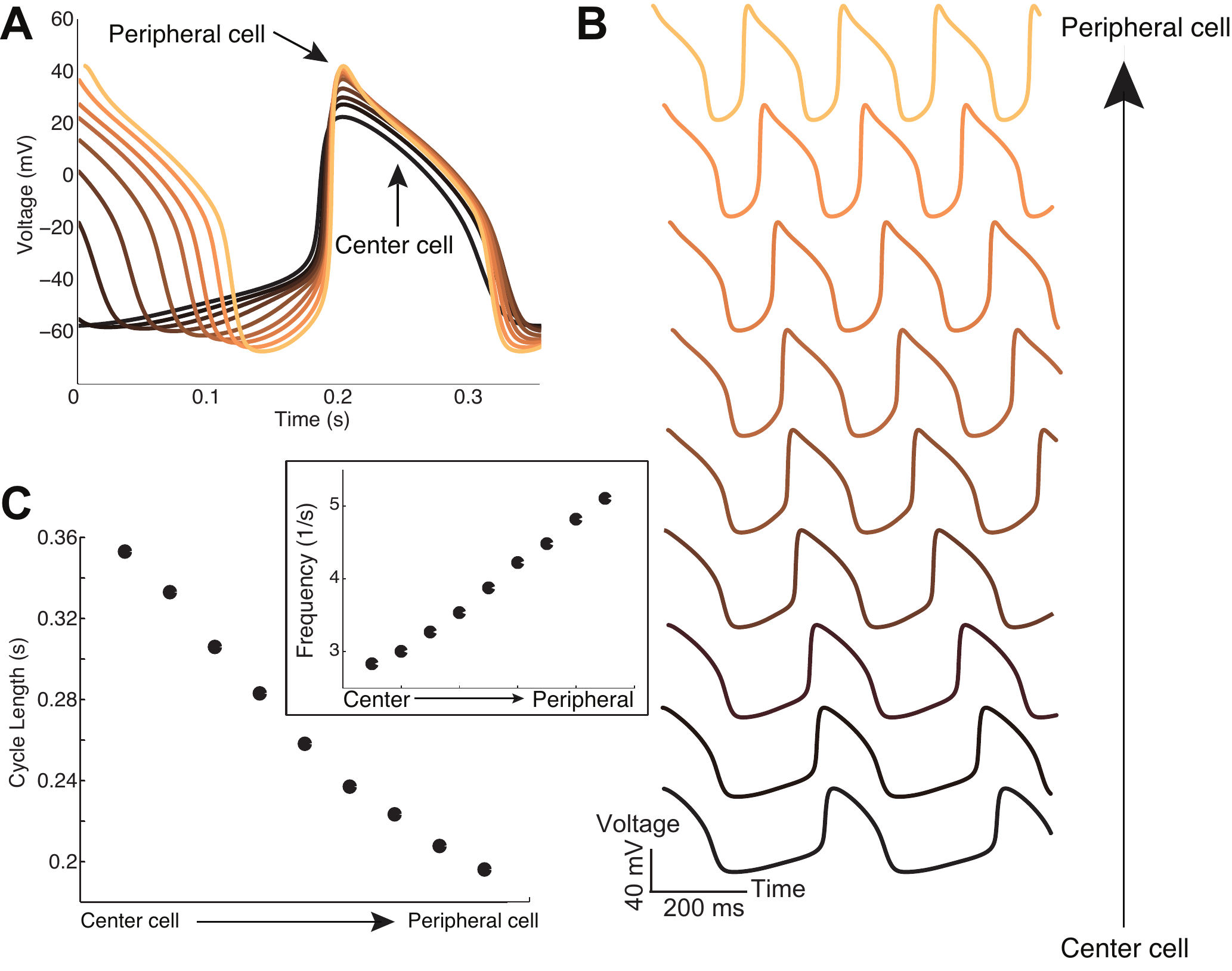}
\caption{The full cardiac pacemaker model, showing all 9 heterogeneous cell types (uncoupled).  
{\bf A}) The spike wave form for all 9 cells, shifted so the peaks are at the same time point to contrast waveforms.  
{\bf B}) Voltage trace of the model over several beats for all 9 cell types.
{\bf C}) The cycle length (in seconds); the inset shows the frequency in 1/s.}
\label{fig:saUncoupled}
\end{center}
\end{figure}

We first developed a population of 
physiologically realistic pacemaker cell models, using the \citet{severi12} model as the baseline
center cell.  This model includes a variety of gating variables, ionic 
currents, and cellular voltage and calcium dynamics.  
By altering 12 parameters, our modified model reproduces the wide-variety of intrinsic frequencies and 
action potential morphology of different pacemaker cell types.  While there is still much debate about the degree of heterogeneity within the SAN cell population, prior experimental work has shown that larger peripheral cells have faster intrinsic frequencies than smaller center cells, 
as well as other electrophysiological properties 
\citep{honjo96,kodama97,opthof87intrinsic}.  
The uncoupled frequency of peripheral cells is almost double ($\approx 1.8 \times$ larger than) the center cells 
(Fig. \ref{fig:saUncoupled}{\bf C} inset), so the 
weak heterogeneity assumption (Eq.\eqref{eqn:H_apprx}) 
does not strictly hold.  
Additionally, action potential morphology is altered, as the voltage trace has a smaller 
minimum and larger maximum and overall 
shorter action potential duration in peripheral cells, compared with center cells 
(Fig. \ref{fig:saUncoupled}{\bf A}), consistent with experimental recordings 
\citep{honjo96,kodama97}.

\begin{figure}[!htb]
\begin{center}
\includegraphics[width=\textwidth]{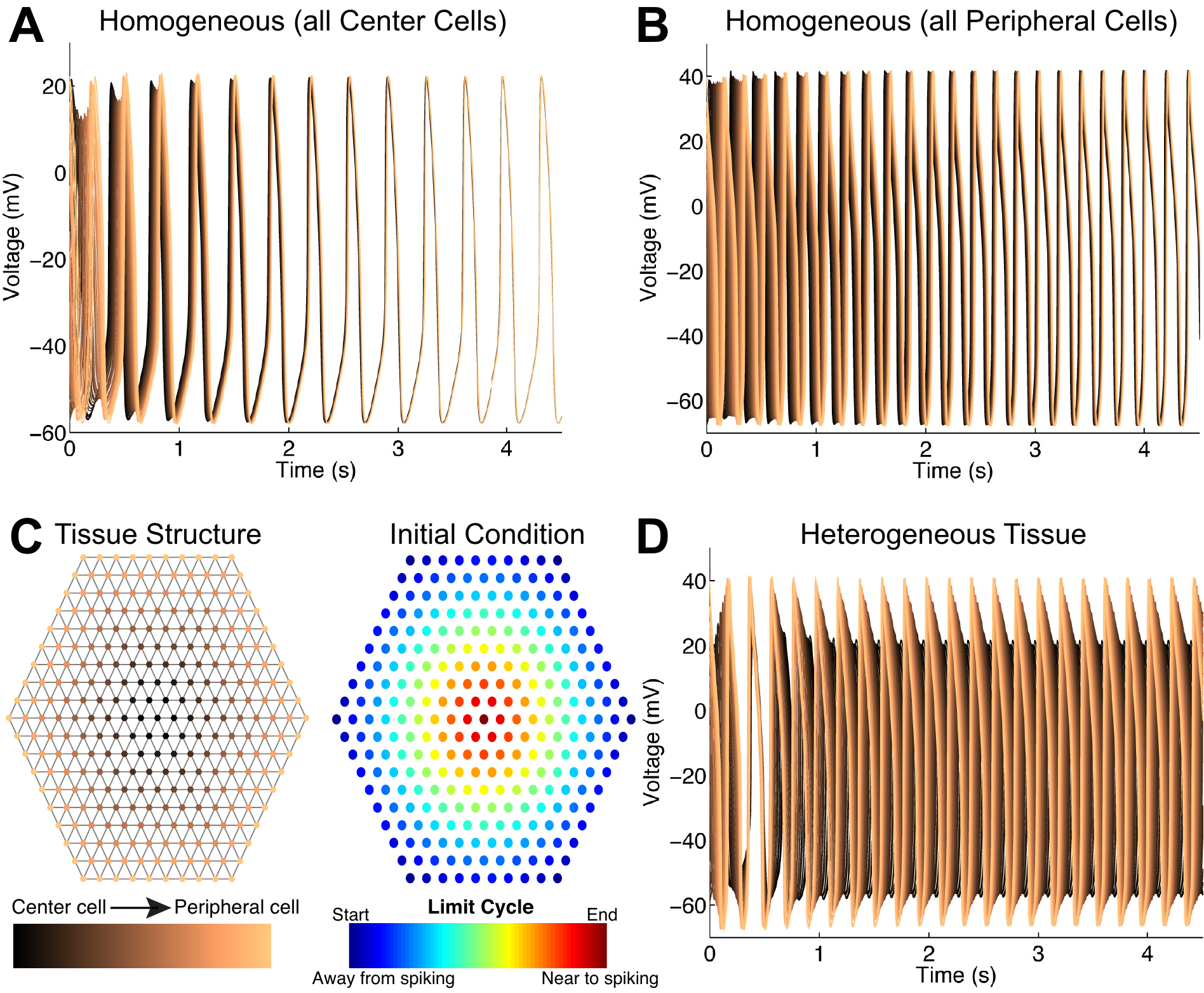}
\caption{Simulation results of the full pacemaker model.  
The color-coding schemes in {\bf A} and {\bf B} do {\bf not} represent intrinsically different cell types, but are chosen to show 
spatial location.  
{\bf A}) With identical cells (all center): after an initial transient of center-to-peripheral traveling wave, 
the system settles to complete synchrony.  
{\bf B}) With identical cells (all peripheral), the transient 
center-to-peripheral wave lasts longer and has 
larger frequency than the center cells in {\bf A}, but eventually settles to complete synchrony.  {\bf C}) 
Left) the architecture is 
nearest neighbor coupling (gray lines connecting dots) used 
for all tissue simulations, the color coding indicates different 
cell types in the heterogeneous tissue used in {\bf D}.  
There are 271 total pacemaker cells on a 
hexagonal grid with all connected cells equidistant apart.  
{\bf C}) Right) the same initial condition for all tissues where 
cells near the center are closer to spiking, and peripheral cells are 
gradually further away from spiking; this sets the tissue to 
transiently have traveling waves originating in the center.  
{\bf D}) In a completely heterogeneous tissue, 
the transient (center-to-peripheral wave) is very brief, after which the system settles to a persistent traveling wave solution that originates in the periphery 
and ends in the center.  In all panels, coupling strength was constant $g_{gap}=0.25\,$nS.
}
\label{fig:saIO}
\end{center}
\end{figure}

We next performed simulations of the SAN tissue models. Simulations of the SAN tissue presented with a variety of spatio-temporal dynamics that depend on coupling strength and cell distribution 
(heterogeneous or homogeneous).  To investigate the role of heterogeneity in the SAN tissue, we simulate a homogeneous tissue comprised 
of either all center (Fig. \ref{fig:saIO}{\bf A}) or all peripheral (Fig. \ref{fig:saIO}{\bf B}) cells, or a heterogeneous tissue with a transition from center to peripheral cell types (Fig. \ref{fig:saIO}{\bf D}). Experimental observations suggest that pacemaking originates in the center of the SAN tissue.  Consistent with these conditions, for all three tissue structures, we set the initial condition of all 31 state variables, such that center 
cells were ``closer" to spiking and 
peripheral cells were ``further" from spiking, in a gradual monotonic way 
(Fig. \ref{fig:saIO}{\bf C} right).  
This resulted in an initial but transient traveling wave that originated in the center for all three tissues. Closer inspection of the voltage traces reveals 
that the transient dynamics have different durations; 
in particular, the transient persists longest in the homogeneous peripheral cell tissue, 
followed by the homogeneous center cell tissue, 
with the heterogeneous tissue having the shortest transient before settling to the steady-state wave solution.  Interestingly, in both homogeneous cases, after an initial transient time period, electrical activity of all of SAN cells became highly synchronized.  However, in the heterogeneous SAN tissue (Fig. \ref{fig:saIO}{\bf C} left), 
a brief transient period is followed by a stable traveling wave, originating in the periphery (Fig. \ref{fig:saIO}{\bf D}), consistent with findings by \citet{oren2010}.  
Note that the coupling strengths are the same for all connected cells and in all three tissues ($g_{gap}=0.25$\,nS).  
Among other tissue attributes, transient times depend on initial conditions; 
here we chose the initial condition primed for central wave generation in all large-scale simulations that did not correspond to the 
steady-state traveling wave for any of the tissue structures.

We did not exhaustively vary the coupling strengths $g_{gap}$ in each tissue structure
because of computational resources, but we did simulate each tissue structure with several sets of gap junction values.  
As coupling strength increases in a given tissue, the transient times 
generally decrease, as one might intuitively expect 
due to stronger interactions between cells (see Fig. \ref{fig:voltWtime1} for the same tissues as Fig. \ref{fig:saIO} but with larger 
$g_{gap}$).  Although this is only apparent by inspection, we will show this is theoretically true in the next 
section with the reduced phase model theory.

Ultrastructure studies of the SAN tissue suggest that there may be an 
increase in gap junctional density from the center to the periphery 
\citep{masson1979plasma,bleeker80,boyett2000sinoatrial}. However, histology 
also illustrates the irregular structure within SAN tissue, consisting of 
discontinuous myofibrils, fat, and fibrous tissue \citep{li15,csepe16}, 
suggesting a wide range of possible cell-cell coupling within the SAN 
tissue.  Therefore, in addition to considering 
tissues with fixed gap junction coupling strength, we also consider tissue 
in which $g_{gap}$ varies depending on 
cell-type and location.  We consider 2 types of tissues: i) a linear 
gradient where peripheral cells have 
$g_{gap}$ 15$\times$'s stronger than center cells (termed {\bf Strong 
Peripheral Gap Junction}, see \citet{oren2010} who also considered this), ii) a linear gradient where center cells 
have $g_{gap}$ 15$\times$'s stronger than peripheral cells (termed {\bf Strong Central Gap Junction}).  
In total, we have three different types of coupling schemes (constant $g_{gap}$ and two 
types of linear gradient $g_{gap}$) that are not directly comparable to each other.  
We chose these tissues to assess the robustness of our observations, and importantly to 
test the robustness of the subsequent reduced phase model theory in the next section.  

Although there are significant differences in the three types of coupling schemes 
(constant $g_{gap}$, Strong Peripheral Gap Junction, 
Strong Central Gap Junction), the following trends hold: 
\begin{itemize}
\item The only steady-state traveling wave solution we observe in 
homogeneous tissue (center or peripheral) is complete synchrony, with our initial conditions.

\item The only observed steady-state traveling wave in the heterogeneous tissue are 
ones that originate in the periphery and end in the center 
(see \citet{kirchhof87} for experimental evidence of peripheral-to-center 
wave in the rabbit SAN when the atrium is removed).  In a given 
tissue, the frequency of this traveling wave 
decreases with gap junction value(s).

\item For the same initial condition (Fig. \ref{fig:saIO}{\bf C} right), 
we observed that: 
the center-to-peripheral traveling wave transient persists longest in the homogeneous peripheral cell-type tissue, 
with the heterogeneous tissue generally having the 
shortest transient times -- there are some exceptions where 
the homogeneous center cell-type tissue has shorter transient times 
than the heterogeneous tissue (especially with larger $g_{gap}$ values), but the transient times for these two tissues 
are very similar in those cases.

\end{itemize}

These results will be substantiated with plots of the 
large-scale simulations 
and characterized with the reduced phase model in 
the next section.  Specific voltage traces 
of the large-scale model for various tissue types, 
coupling schemes $g_{gap}$ 
values, etc., 
are shown in Figs. \ref{fig:voltWtime1}--\ref{fig:voltWtime3} and Figs. S3--S5 (S1Text.pdf).  

\subsection{Reduced Phase Model Analysis of Traveling Waves}

To more clearly understand the observations in the large-scale simulations, 
we use a reduced phase model framework.  This 
will not only make mathematical analysis of the tissue dynamics feasible, 
but will also provide a framework for 
future investigations of large-scale tissue models of SAN tissue.  
The subsequent analysis relies on computing various functions from the \underline{individual Severi} 
cell models, but did not rely on the computationally expensive step of simulating the large-scale tissue model.

\begin{figure}[!htb]
\begin{center}
\includegraphics[width=\textwidth]{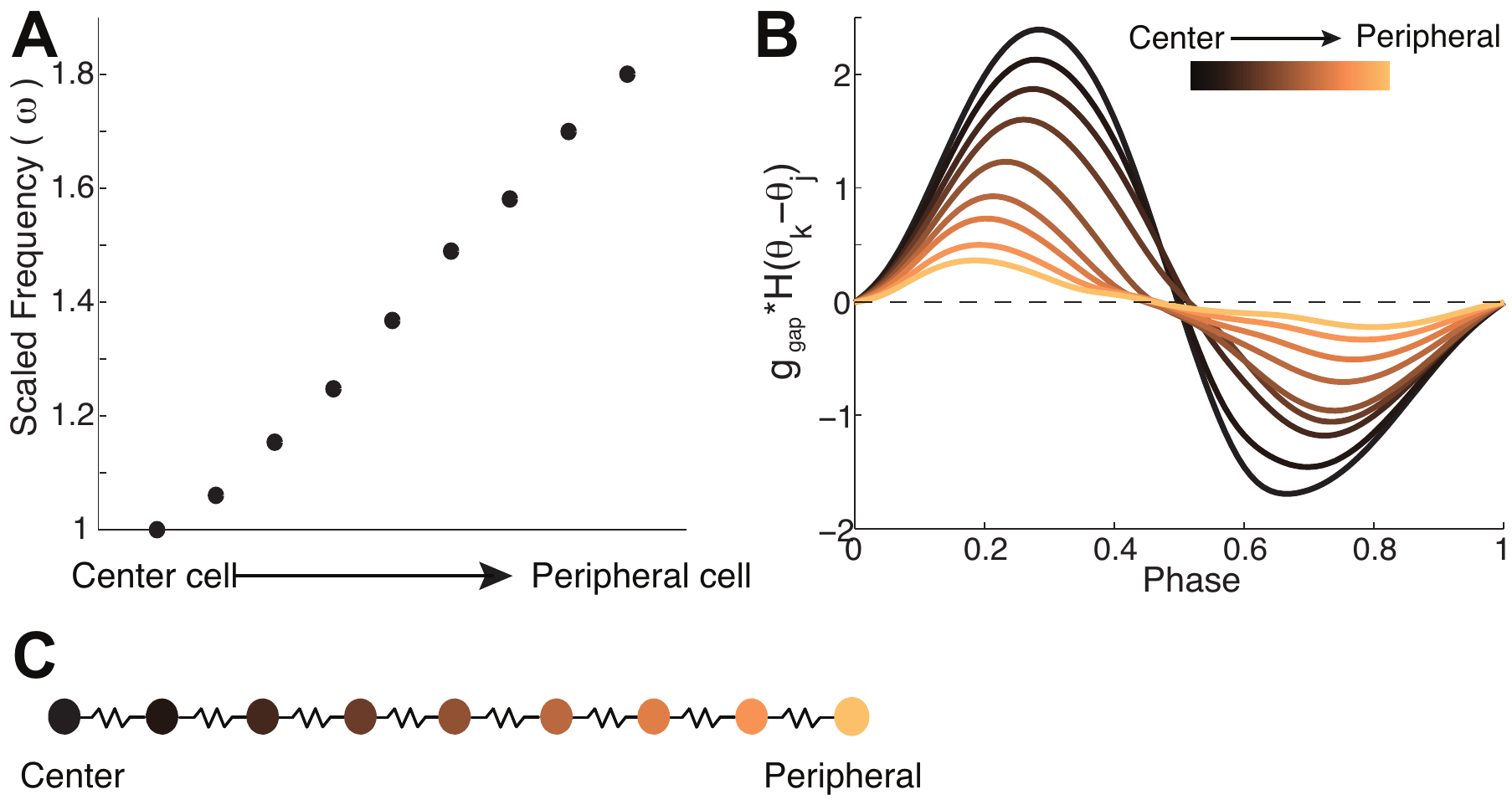}
\caption{The phase oscillator model approximation to the large-scale 
Severi pacemaker model.  
{\bf A}) The intrinsic frequencies were obtained from the full model and scaled by the largest period.  
{\bf B}) The interaction functions $H_{j,k}$ are approximated with $H_j$ 
for $k=j-1$ and $k=j+1$; see Eq~(\ref{eqn:H_apprx2}).  We set $g_{gap}=1$\,nS for illustration purposes.  
{\bf C}) The 1D chain approximation to the full 2D model.}
\label{fig:phase1}
\end{center}
\end{figure}

We use the phase reduced models (see {\bf Methods}) to approximate the behavior of the large-scale SAN tissue, 
where each cell type has an equation of the form:
\begin{equation}\label{eqn:redModelFull}
	\frac{d\theta_j}{dt} = \omega_j + \sum_k H_{j,k}(\theta_k-\theta_j) 
\end{equation}
The frequencies $\omega_j$ are all 
scaled by the slowest intrinsic frequency (often the center cell with $T=0.353$\,s), such that 
in the heterogeneous tissue, the center cells have $\omega=1$ 
and peripheral cells $\omega\approx 1.8$ (see Fig. \ref{fig:phase1}{\bf A} for the full range 
of possible $\omega$ values and Fig. \ref{fig:phase1}{\bf B} for unscaled 
versions of $H_{j,k}$).  
Note that the  intrinsic period of the peripheral cell is $T=0.196$\,s.  

The reduced phase model consists of a 1D chain of $9$ phase oscillators with nearest neighbor 
coupling: $H_{j,k}\equiv 0$ unless $k=j-1$ or $k=j+1$ (see Fig. \ref{fig:phase1}{\bf C}), such that Eq.~(\ref{eqn:redModelFull}) for 
this 1D chain is re-written as:
\begin{eqnarray}
    \frac{d\theta_1}{dt} &=& \omega_1 + H_{1,2}(\theta_2-\theta_1) \nonumber \\
    \frac{d\theta_j}{dt} &=& \omega_j + H_{j,j-1}(\theta_{j-1}-\theta_j)+H_{j,j+1}(\theta_{j+1}-\theta_j), \hbox{  for }j= 2,\dots, 8 \nonumber \\
    \frac{d\theta_{9}}{dt} &=& \omega_{9} + H_{9,8}(\theta_{8}-\theta_{9}) \label{eqn:redModel}
\end{eqnarray}

This model is amenable to analysis of traveling wave dynamics \citep{kopell86,ermentroutBook,keenerSneyd}.  
Traveling wave solutions to this system are of the form
\begin{equation}\label{eqn:twForm}
	\theta_j(t) = \Omega t + \xi_j
\end{equation}
where $\Omega$ is the ensemble (scaled) frequency of the coupled network and $\xi_j$ 
represents the phase lags of each cell type (i.e., $\xi_j=0$ for all $j$ is complete 
synchrony).  Substituting 
Eq. (\ref{eqn:twForm}) into Eq. (\ref{eqn:redModel}) gives:
\begin{equation}\label{eqn:exist_tw}
	\Omega = \omega_j + \sum_k H_{j,k}(\xi_k-\xi_j), \hbox{ for }j=1,\dots,9.
\end{equation}
The system of 9 nonlinear equations and 10 unknowns ($\Omega$, $\xi_j$) 
can be reduced to 9 unknowns by exploiting the periodicity of $\xi_j\in[0,1)$, 
and setting $\xi_1=0$.  Thus, if $\xi_j-\xi_{j+1}>0$ for $j=2,\dots,8$ the traveling wave 
originates at the center ($j=1$), and if $\xi_{j+1}-\xi_j>0$ the traveling wave originates 
in the periphery (see \citet{ermentroutBook} for similar exposition) 
\footnote{The exception is if the traveling wave solution has 
very fast frequency and the next wave starts before the current one ends, 
in which case the sign of $\xi_{j+1}-\xi_j$ is different for a $j$, 
see Fig.\ref{fig:lngPhase}{\bf B} in yellow and 
Fig.\ref{fig:nonTW}{\bf D,E,F}.}.  

\underline{Traveling wave solutions to the reduced phase model 
system exist} (numerically) whenever 
the variables $(\Omega,\vec{\xi})$ satisfy the Eqs.~(\ref{eqn:twForm})--(\ref{eqn:exist_tw}).  
For each of the coupling schemes and all three tissue types (2 homogeneous, uniform heterogeneous), 
we found several sets of $g_{gap}$ values where traveling wave solutions 
numerically existed.  
Solutions to Eqs.~(\ref{eqn:twForm})--(\ref{eqn:exist_tw}) 
were found using {\tt fmincon} in MATLAB with high precision (agreement of at least 3 decimal places in each equation). 

\begin{figure}[!htb]
\begin{center}
\includegraphics[width=.85\textwidth]{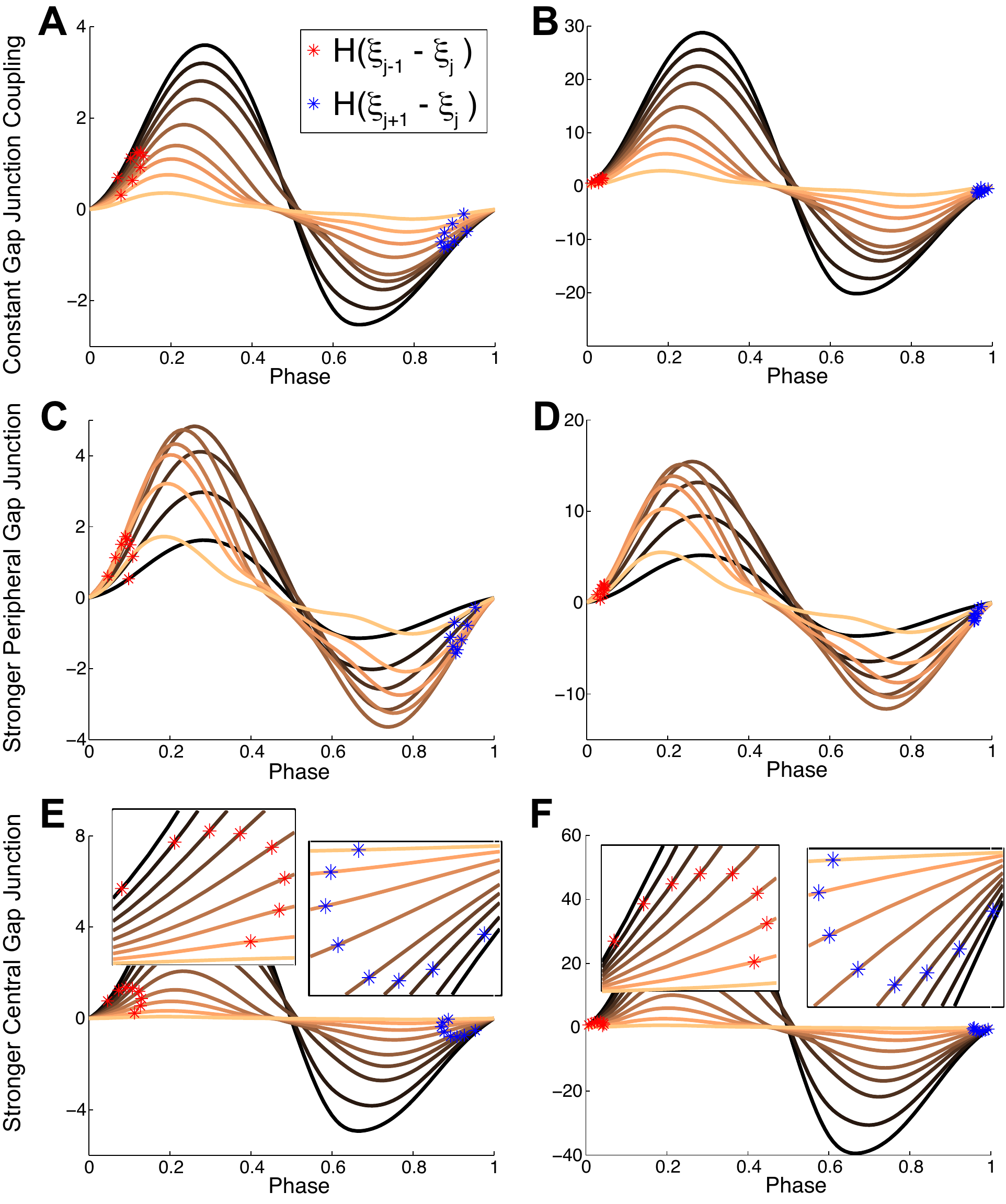}
\caption{Existence and stability of traveling waves for heterogeneous tissues.  
The following plots show the $H$-functions for a specific gap junction coupling strengths, along with calculated quantities $H_{j,k}(\xi_{j-1}-\xi_j)$ 
and $H_{j,k}(\xi_{j-1}-\xi_j)$.  A sufficient condition for stability of the waves relies, in part, on $\frac{dH}{d\theta}(\xi_{j-1}-\xi_j)>0$ and 
$\frac{dH}{d\theta}(\xi_{j+1}-\xi_j)>0$ (see main text), which holds for 
all tissues here.  In all the regimes we explored with heterogeneous cells, 
traveling waves originating in peripheral cells exist and are stable.  
{\bf A}) For constant gap junction coupling, with $g_{gap}=0.25$\,nS in the full large-scale tissue.  
{\bf B}) Same as {\bf A} but with larger $g_{gap}=2$\,nS.  
{\bf C}) Gap junction strength $g_{gap}$ varying monotonically from center to peripheral, with largest value 
at the peripheral cells ($g_{gap}=1.25$\,nS) 
and smallest value at center cells ($g_{gap}=1.25/15$\,nS).  {\bf D}) 
Same as {\bf C} but with much larger coupling, $g_{gap}=4$\,nS at peripheral cells 
and $g_{gap}=4/15$\,nS at center cells.
{\bf E}) Gap junction strength varying monotonically, with largest at the center cells ($g_{gap}=0.5$\,nS) and smallest at peripheral cells ($g_{gap}=0.5/15$\,nS).  
The 2 insets are zoomed-in pictures to convincingly show that the stars are where $H$ is increasing (i.e., $\frac{dH}{d\theta}>0$).  
{\bf F}) Same as {\bf E} but with much larger coupling, $g_{gap}=4$\,nS at center cells and $g_{gap}=4/15$\,nS at peripheral cells.
}
\label{fig:stability}
\end{center}
\end{figure}

To address the stability of traveling wave solutions when they exist, we rely 
on a theorem by \citet{kopell86,ermentrout92} which 
provides sufficient conditions for the stability of traveling wave solutions.  
The theorem re-stated for these reduced phase model networks is:

\begin{theorem} The traveling wave solution $(\Omega,\vec{\xi})$ that satisfies 
Eq.~(\ref{eqn:exist_tw}) is stable if the coupled network is \textit{irreducible} and 
$H_{j,k}'\left(\xi_k-\xi_j\right)\geq 0$.
\end{theorem}

The proof is technical and can be found in \citet{kopell86,ermentrout92}.  
We prove the stability of traveling waves by applying this theorem to 
a range of parameter sets in the 3 tissue types and coupling regimes.  
Note that the chain of oscillators (Fig. \ref{fig:phase1}{\bf C}) 
is irreducible since every oscillator is connected, so the only condition 
to check is $H_{j,k}'\left(\xi_k-\xi_j\right)\geq 0$.  
In correspondence with the large-scale tissue model, recall that 
we consider: i) constant gap junction coupling, and gradient gap junction 
coupling with ii) a linear gradient where peripheral cells have 
$g_{gap}$ 15$\times$'s stronger than center cells (termed {\bf Strong Peripheral Gap Junction}), 
and iii) a linear gradient where center cells 
have $g_{gap}$ 15$\times$'s stronger than peripheral cells (termed {\bf Strong Central Gap Junction}).

Synchrony is the only periodic steady state for the 
homogeneous tissues 
(all center cells or all peripheral) in the phase models; 
we explain why this state is always stable.  
With complete synchrony, $\xi_j=0$ for all $j$ in Eq.~(\ref{eqn:twForm}), so we only have to 
check that $H'_{j,k}(0)\geq 0$ according to the theorem.  This is easily verified in Fig. \ref{fig:phase1}{\bf B}, and 
holds for any $g_{gap}>0$ because it simply scales the the curve, 
as well as any other homogeneous tissue with any one of the 9 cell types.  
Thus, complete synchrony is always a stable state in 
the homogeneous tissues we consider.

In the heterogeneous tissues, the traveling wave solutions consist of 
$\xi_j>0$, so we have to verify that $H'_{j,k}(\xi_k-\xi_j) \geq 0$ 
numerically to show that the solution is stable.  We initially 
considered 15 total parameter sets for the large-scale 
heterogeneous tissues; we  discovered traveling wave solutions existed in 14 
(5 with constant $g_{gap}$ coupling, 4 with strong peripheral, 5 with strong 
central), 
and in all these cases, the traveling waves are stable. We discuss the parameter set without a traveling wave solution below. Fig. \ref{fig:stability} shows the smallest (left column) 
and largest coupling values (right column) considered, 
with each coupling schemes organized by row.  
Notice that in all instances, $H'_{j,k}>0$, indicating that the traveling 
wave is stable.  Stability of 
all other heterogeneous parameter sets are shown in Fig. S1 
(S1Text.pdf).

\begin{figure}[!htb]
\begin{center}
\includegraphics[width=.4\textwidth]{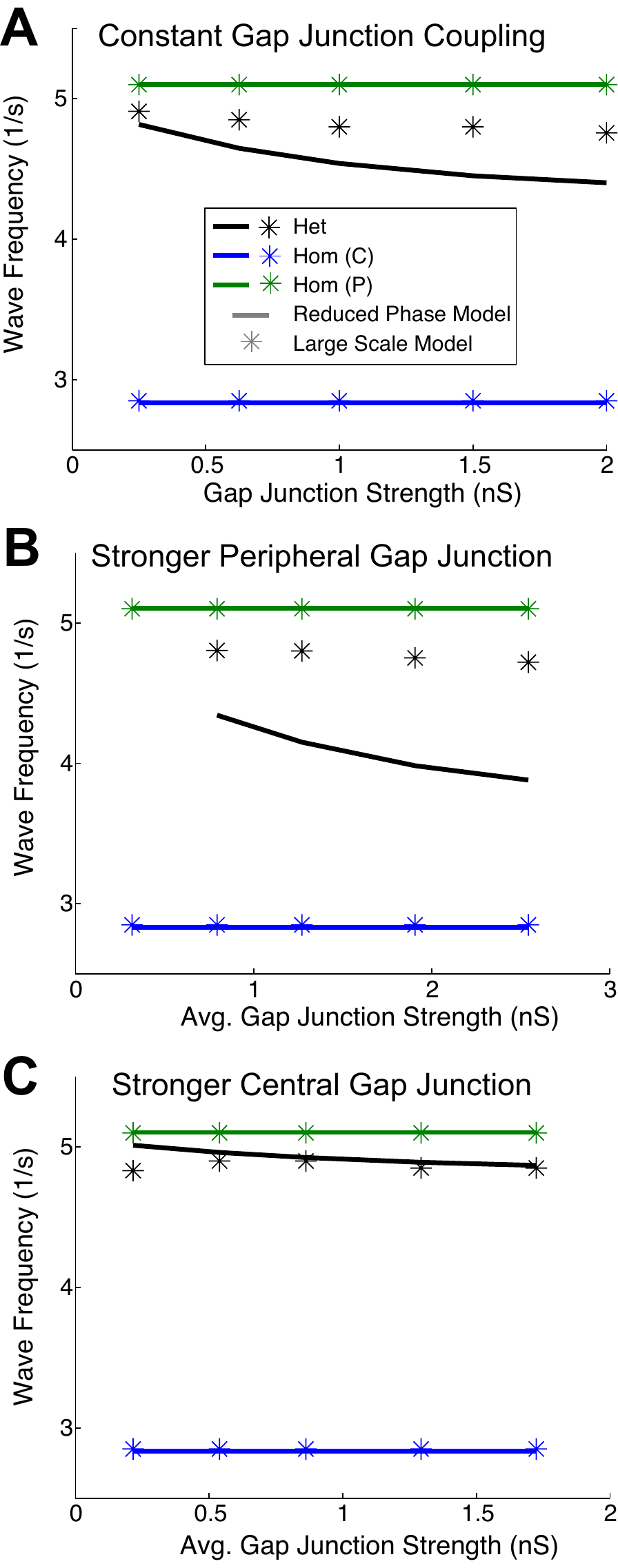}
\caption{Reduced phase model theory 
captures the traveling wave frequency in the large-scale model.  
The ensemble traveling wave frequency in the reduced phase model ($\Omega$ in Eq.~(\ref{eqn:twForm}) scaled to 1/sec) plotted 
with solid lines, is compared to the traveling wave frequency of the full large-scale coupled tissue model (stars).  
Here we only consider the reduced phase models that correspond to the parameter sets we initially considered in the large-scale models.  
In all panels, the periodic steady state for both homogeneous Center (blue) and Peripheral (green) cell tissues 
is complete synchrony; 
in the heterogeneous tissue (black), the wave originates in the periphery and terminates somewhere near the center.  
{\bf A}) For constant gap junction coupling, varying $g_{gap}$ from 0.25\,nS to 2\,nS in the full large-scale tissue.  
{\bf B}) Gap junction strength $g_{gap}$ varying monotonically from center (weakest) to peripheral (strongest); 
the x-axis shows the average $g_{gap}$ among all connected 
cells for a given tissue configuration.  
In the heterogeneous tissue, 
a traveling wave solution does not exist for smaller $g_{gap}$.  
{\bf C}) Gap junction strength $g_{gap}$ varying monotonically from center (strongest) to peripheral (weakest); x-axis has the 
same convention as {\bf B}.
}
\label{fig:cfFreqs}
\end{center}
\end{figure}

The coupled reduced phase model can also 
qualitatively reproduce the key property of tissue frequency 
from the large-scale model in 
both uniformly coupled homogeneous (either all center or all peripheral) or heterogeneous population of SAN cells.  This is demonstrated 
with a variety of coupling schemes and values: with constant gap junction (Fig. \ref{fig:cfFreqs}{\bf A}), 
strong peripheral gap junction (Fig. \ref{fig:cfFreqs}{\bf B}), strong central gap junction (Fig. \ref{fig:cfFreqs}{\bf C}).  
The solid lines in Fig. \ref{fig:cfFreqs} are the $\Omega$ 
from the reduced phase model, using Eq.~(\ref{eqn:twForm}) but 
scaled by the smallest intrinsic frequency to have units of 
(1/sec), while the tissue frequencies from the large-scale model 
(stars) are calculated by taking the average of the peak frequencies of the 
power spectrums of the voltage (in the last 10\,s out of 20\,s of 
biological time) for all 271 pacemaker cells.  

Although the matches between the curves (phase model) and 
stars (large-scale simulations) in Figure \ref{fig:cfFreqs} 
are not quantitatively accurate for heterogeneous tissues, 
the qualitative trend that wave frequency decreases with gap junction 
strength is captured.  The comparisons deviate most with 
strong peripheral (gradient coupling strength, 
Fig. \ref{fig:cfFreqs}{\bf B}), and are 
most accurate with strong central (gradient coupling 
strength, Fig. \ref{fig:cfFreqs}{\bf C}); the match with constant gap 
junction coupling in Figure \ref{fig:cfFreqs}{\bf A} is 
better than {\bf B} and worse than {\bf C}.  This is all likely due 
to the violation (or conformity) of the 
weak heterogeneity assumption in the reduced phase model.  
This assumption is severely violated 
with stronger peripheral gap junction because the 
larger number of peripheral cells with the hexagonal grid 
(Fig. \ref{fig:saIO}{\bf C} left) \textit{combined} with larger $g_{gap}$ 
results in a phase model with 
large heterogeneity from cell-type to cell-type 
(cell-types are grouped and averaged in the reduced phase model).  
Recall that the intrinsic frequency difference varies greatly by almost a 
factor of 2 ($\approx 1.8 \times$).  
Conversely, with stronger central gap junction, the larger 
number of peripheral cells is counter balanced by weaker $g_{gap}$, 
resulting in weaker heterogeneity from cell-type to cell-type.  
Note that the deviations in wave frequency are 
less than 1\,Hz but that the intrinsic frequencies differ by 2.27\,Hz. 

Finally, since the dynamics of the large-scale pacemaker tissue model only 
transiently reproduce the physiologically-observed center-to-peripheral 
wave 
pattern, one possible mechanism explaining this response is that autonomic 
regulation may regularly reset ``initial conditions'' to drive the 
center-to-peripheral wave.  Therefore, we sought to quantify the characteristics 
of the transient dynamics, before 
the coupled system settles to a stable traveling wave, as well as the 
agreement between the large-scale simulations and reduced model.   A 
perturbation 
$\vec{\eta}(t)$ to a traveling wave solution  
$\vec{\theta}(t)=\Omega t+\vec{\xi}$ is:
\begin{equation}\label{eqn:perturb}
	\theta_j(t)=\Omega t + \xi_j + \eta_j(t);
\end{equation}
the transient dynamics can be approximated by assuming 
perturbations are small $\Big( \eta_j=O(\varepsilon)$ for all $j \Big)$; 
see {\bf Methods} section for further details.  
This equation in matrix-vector form is (see equation 
(\ref{eqn:Adefn})):
$$ \frac{d\vec{\eta}}{dt} = A \vec{\eta} $$

When a traveling wave solution is stable, 
all but one of the eigenvalues of $A$ 
have negative real part; there is a 
a distinct zero eigenvalue $\lambda_0=0$ with eigenvector $\vec{1}$ that corresponds to a constant shift in 
the traveling wave solution -- a constant shift in all variables is a perturbation that does not decay 
because the result is the same traveling wave solution.  

\begin{figure}[!htb]
\begin{center}
\includegraphics[width=.8\textwidth]{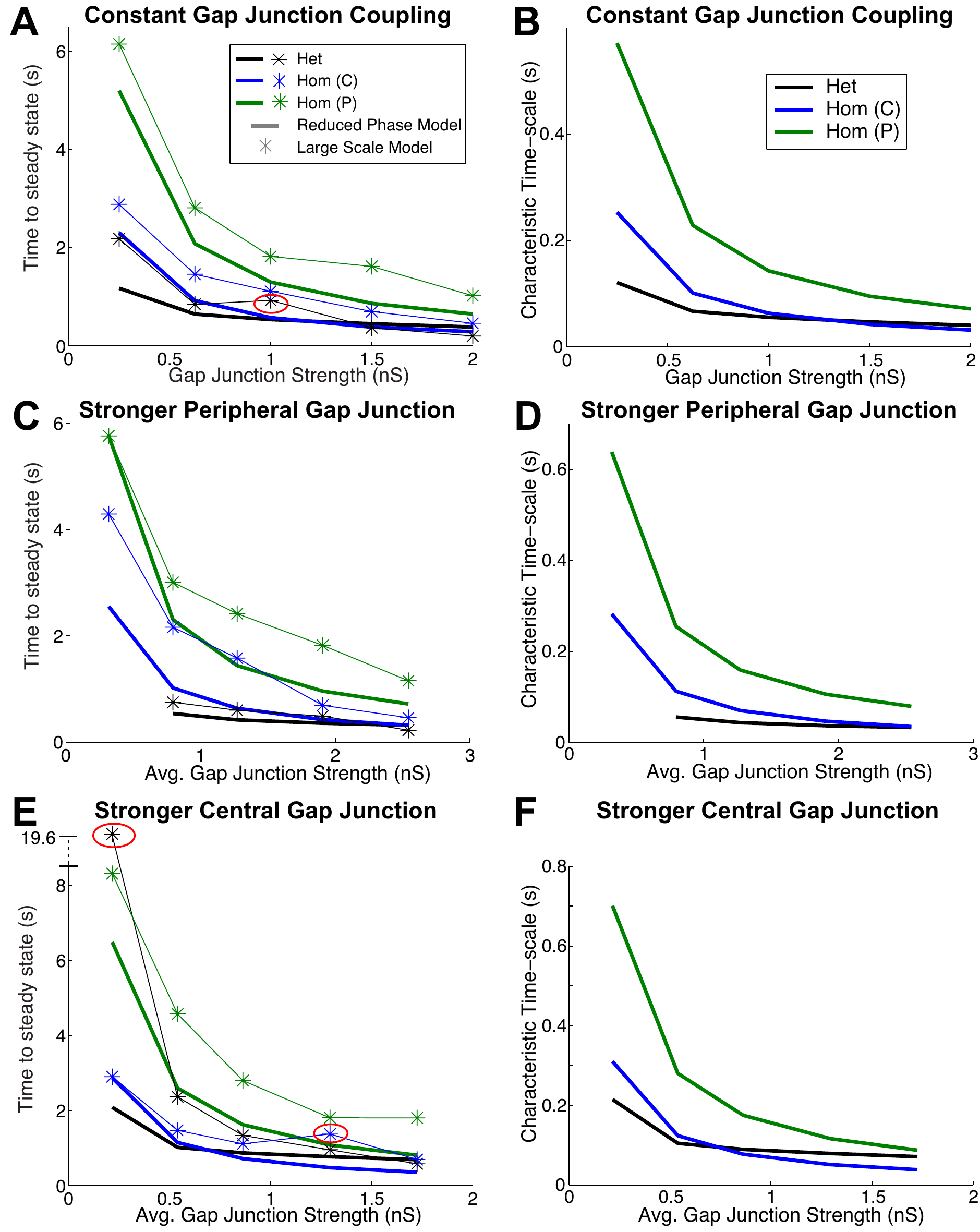}
\caption{Reduced phase model theory captures the 
transient times before settling to steady-state traveling waves.  
Left Column: for a given tissue configuration with initial 
condition primed to center-to-peripheral traveling waves 
(Fig. \ref{fig:saIO}{\bf C} right), the transient times 
from largest to smallest are generally: homogeneous peripheral cell tissue (green), homogeneous center cell tissue (blue), 
heterogeneous tissue (black).  As coupling strength increases, transient times decrease and 
the differences between blue and black are diminished.  
Although the match between large-scale tissue (stars) and reduced phase model 
(solid curves) is not always precise, the qualitative trends are captured. 
The right column shows the characteristic time-scale ($(\vert \lambda_1 \vert)^{-1}$, where $\lambda_1$ is the largest 
negative eigenvalue of $A$; see Eq.~(\ref{eqn:etaMatr}) but rescaled to seconds), which is a measure of transient time that does {\bf not} 
depend on initial condition or numerical tolerance.  
{\bf A, B}) For constant gap junction coupling, varying $g_{gap}$ from 0.25\,nS to 2\,nS in the full large-scale tissue.  
{\bf C, D}) Gap junction strength $g_{gap}$ varying monotonically from center (weakest) to peripheral (strongest); 
the x-axis shows the average $g_{gap}$ among all connected 
cells for a given tissue configuration.  
{\bf E, F}) Gap junction strength $g_{gap}$ varying monotonically from center (strongest) to peripheral (weakest); x-axis has the 
same convention as {\bf C, D}.  
The 3 red ovals are when the 
effects from the boundaries of the hexagonal grid persist 
so that the 1D chain is 
is a bad approximation -- see Supplementary Material 
(from top to bottom: vidx\_Het\_g3\_s3.mp4,  vidx\_Het\_g2\_s2.mp4, 
vidx\_HomC\_g33\_s3.mp4 in S2.zip).  
}
\label{fig:trTime}
\end{center}
\end{figure}

\begin{figure}
\centering
\includegraphics[width=\textwidth]{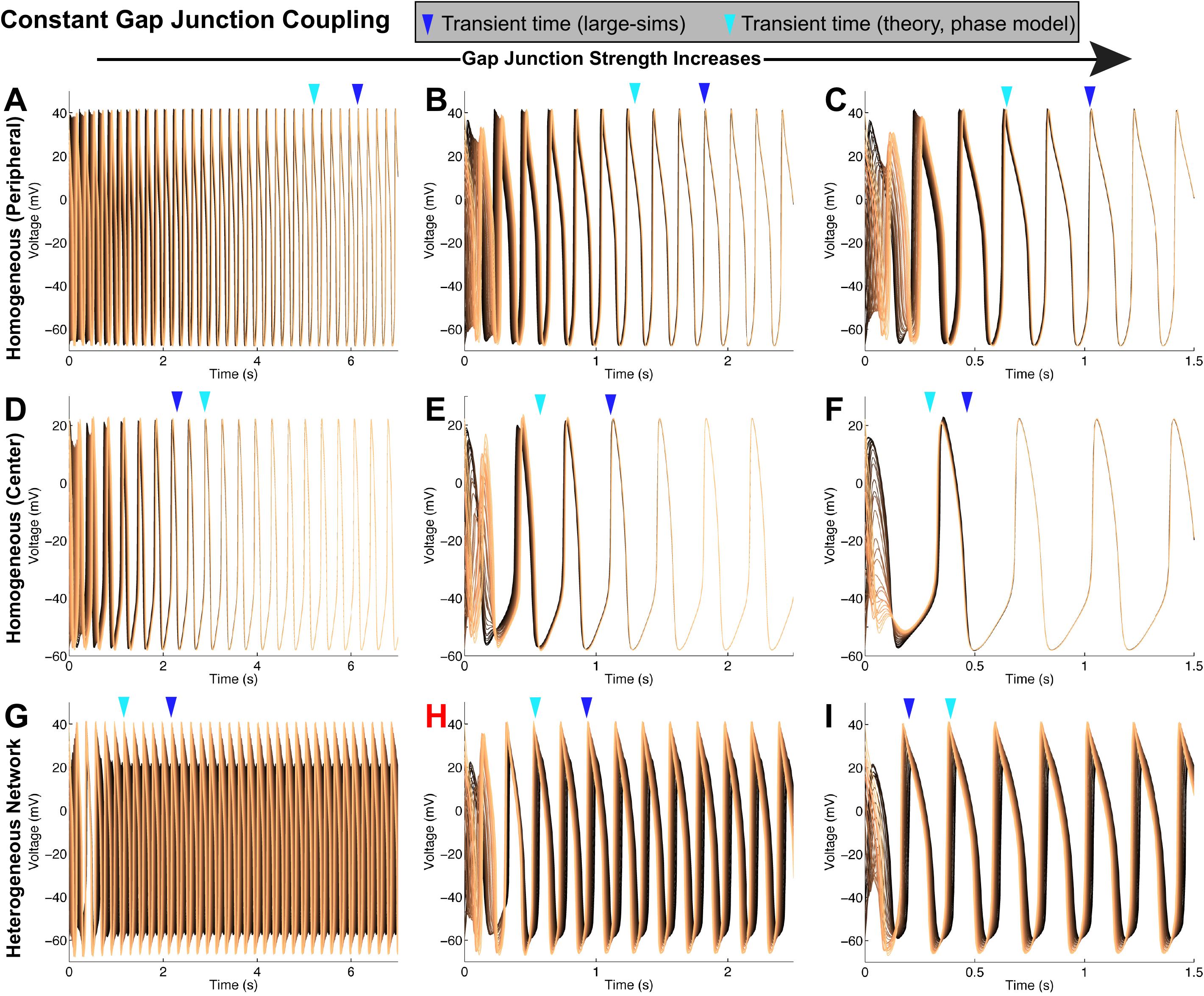}
\caption{\label{fig:voltWtime1} Voltage trajectories in large-scale model plotted with the theoretical transient time (reduced phase model) and 
large-scale transient time: with constant gap junction coupling.  
By inspection, the transient times from the reduced phase model 
(cyan arrow) and large-scale model (blue arrow) 
are reasonable.  
Each column has different coupling strengths, each row is a particular type of tissue.  
Left column ({\bf A}, {\bf D}, {\bf G}) is with $g_{gap}=0.25$\,nS.  
Middle column ({\bf B}, {\bf E}, {\bf H}) 
is with $g_{gap}=1$\,nS.  
Right column ({\bf C}, {\bf F}, {\bf I}) is with $g_{gap}=2$\,nS.  
{\bf A})--{\bf C}) The homogeneous tissue of peripheral cells.  
{\bf D})--{\bf F}) The homogeneous tissue of center cells.  
{\bf G})--{\bf I}) The heterogeneous tissue consisting of 9 different cell types.  
In each column, the transient generally last longest for homogeneous peripheral 
cell tissue (top row), following by homogeneous center cells, with the heterogeneous tissue having the shortest transient.  The label {\bf H} is red to correspond with the 
red oval in Figure \ref{fig:trTime}{\bf A}.
} 
\end{figure}

\begin{figure}
\centering
\includegraphics[width=\textwidth]{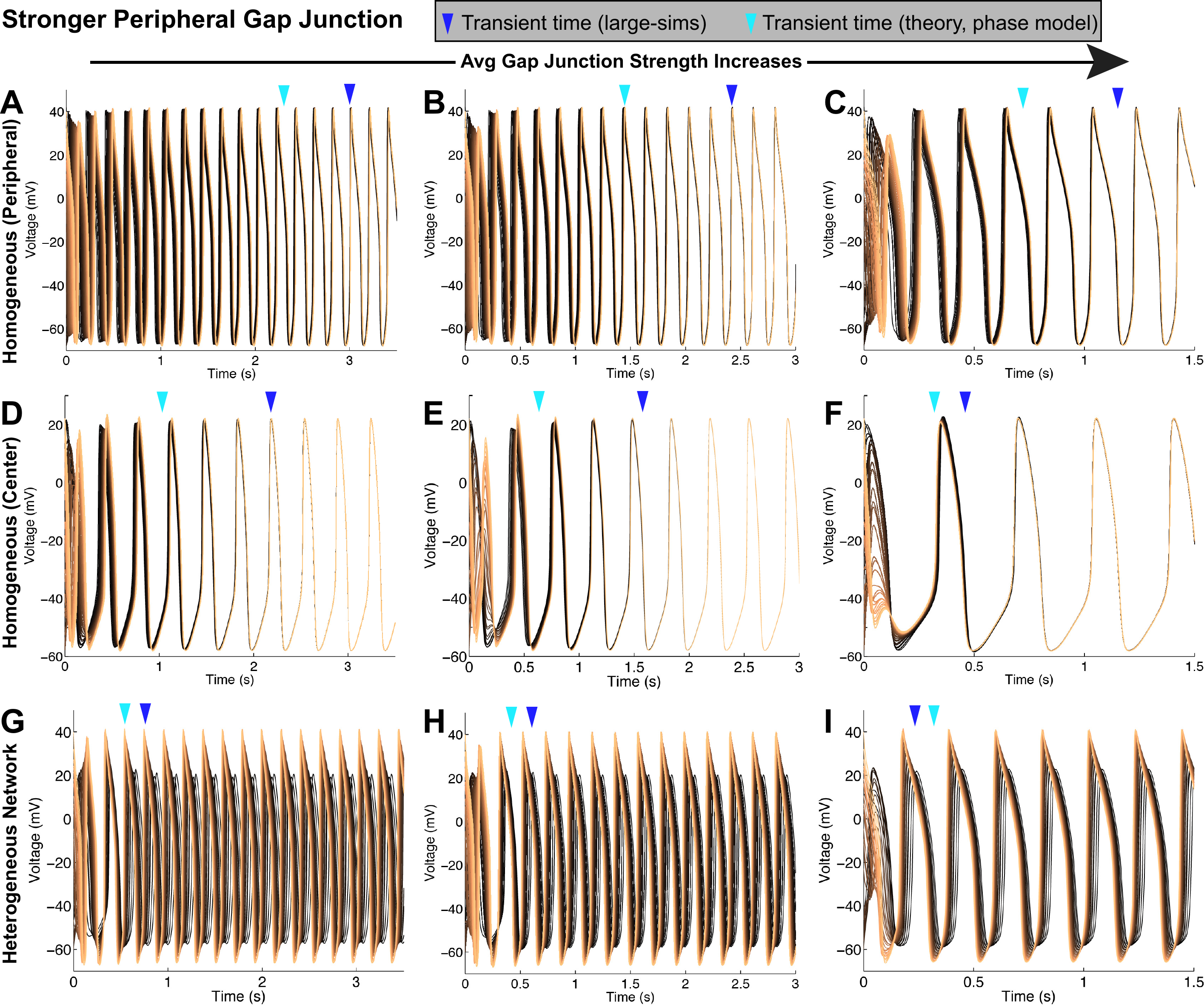}
\caption{\label{fig:voltWtime2} Voltage trajectories in large-scale model plotted with the theoretical transient time (reduced phase model) and 
large-scale transient time: with strong peripheral coupling.  
Again by inspection, the transient times from the reduced phase model 
(cyan arrow) and large-scale model (blue arrow) 
are reasonable.  
Each column has different coupling strengths, each row is a particular type of tissue (same format as Fig. \ref{fig:voltWtime1}).  
Left column ({\bf A}, {\bf D}, {\bf G}) is with $\frac{0.625}{15}\text{\,nS}\leq g_{gap}\leq 0.625$\,nS.  
Middle column ({\bf B}, {\bf E}, {\bf H}) 
is with $\frac{2}{15}\text{\,nS} \leq g_{gap} \leq 2$\,nS.  
Right column ({\bf C}, {\bf F}, {\bf I}) 
is with $\frac{4}{15}\text{\,nS}\leq g_{gap} \leq 4$\,nS.  
In each column, the transient generally last longest for homogeneous peripheral 
cell tissues (top row), following by homogeneous center cells, with the heterogeneous tissues having the shortest transient.  
} 
\end{figure}

\begin{figure}
\centering
\includegraphics[width=\textwidth]{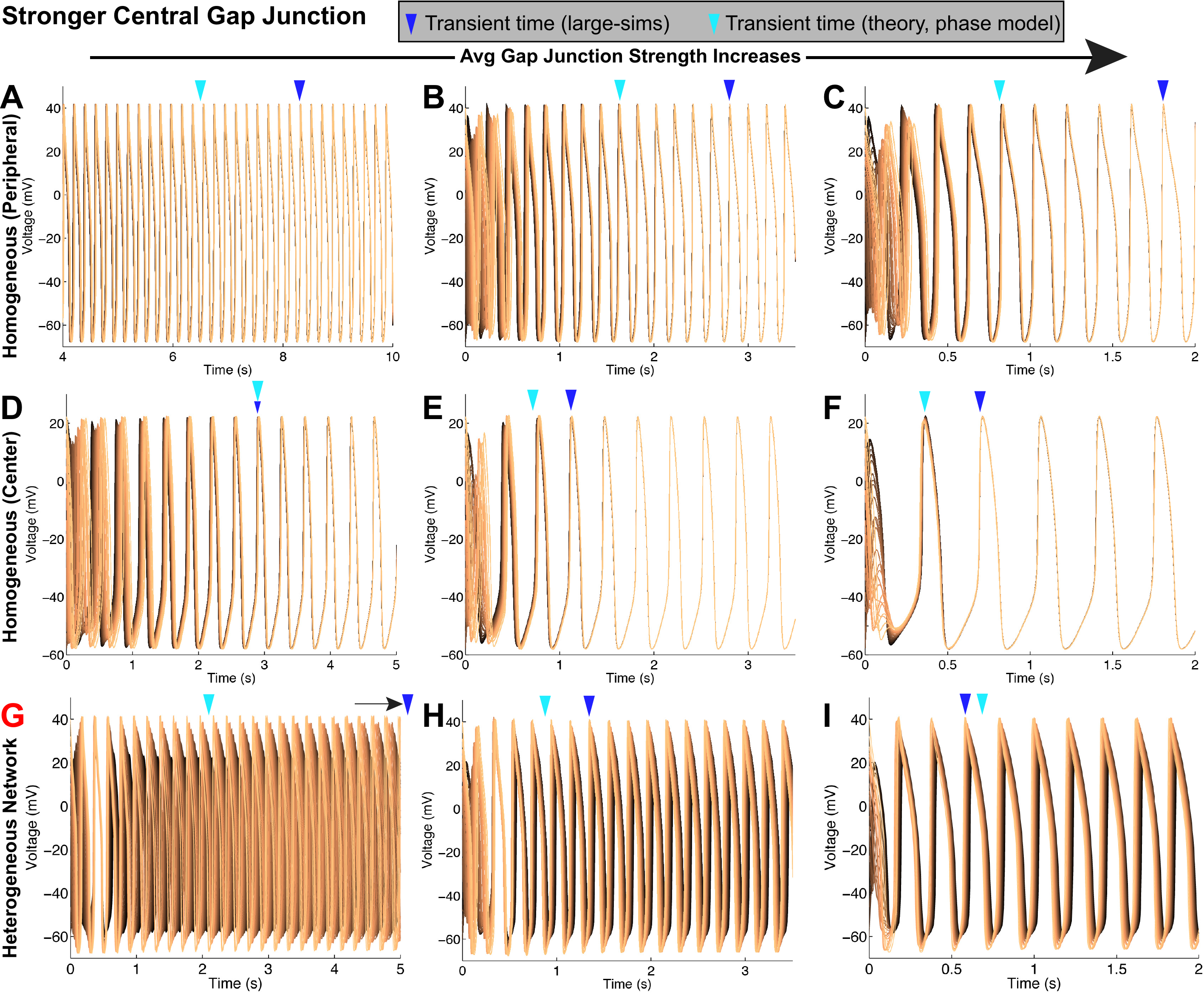}
\caption{\label{fig:voltWtime3}  Voltage trajectories in large-scale model plotted with the theoretical transient time (reduced phase model) and 
large-scale transient time: with strong central coupling.  
Again by inspection, the transient times from the reduced phase model 
(cyan arrow) and large-scale model (blue arrow) 
are reasonable.  
Each column has different coupling strengths, each row is a particular type of tissue (same format as Fig. \ref{fig:voltWtime1}).  
Left column ({\bf A}, {\bf D}, {\bf G}) is with $\frac{0.625}{15}\text{\,nS} \leq g_{gap} \leq 0.625$\,nS.  
Middle column ({\bf B}, {\bf E}, {\bf H}) 
is with $\frac{2}{15}\text{\,nS}\leq g_{gap} \leq 2$\,nS.  
Right column ({\bf C}, {\bf F}, {\bf I}) 
is with $\frac{4}{15}\text{\,nS} \leq g_{gap} \leq 4$\,nS.  
In each column, the transient generally last longest for homogeneous peripheral 
cell tissue (top row), following by homogeneous center cells, with the heterogeneous tissue having the shortest transient.  The label {\bf G} is red to correspond with the top red oval in Figure \ref{fig:trTime}{\bf E}.
} 
\end{figure}

Recall that the vector of perturbations can be 
approximated by using an eigenvector expansion 
(see {\bf Methods} section).  
Let the eigenvectors of $A$ be $\vec{\psi}_j$ with eigenvalues $\lambda_j$: $A\vec{\psi}_j = \lambda_j\vec{\psi}_j$.  We 
use the initial condition from the large-scale 
(Fig. \ref{fig:saIO}{\bf C} left) mapped to the phase variable; 
this is well-approximated by $\vec{\eta}_0(j)=1-j*\frac{1}{10}$ 
(traveling wave starting at center) but we use the actual mapped phase value.  
The vector of perturbations $\vec{\eta}$ is approximated via: 
$$\vec{\eta}(t)=  (\vec{\psi}_1,\vec{\eta_0}) e^{\lambda_1 t} \vec{\psi_1}$$
where $\lambda_1$ is the eigenvalue with large nonzero real part, 
and $(\vec{v},\vec{w})$ is the usual inner product.  
In the Supplementary Material (S1Text.pdf), we show 
that this solution is numerically the same as the full eigenvector expansion (see Fig. S2 in S1Text.pdf).  
The times to the steady-state traveling wave solutions are approximated by determining when the perturbations 
decay to a specified tolerance: $ \| \vec{\eta}(t) \| < \epsilon$.  
In all figures (\ref{fig:trTime}{\bf A, C, E}, \ref{fig:voltWtime1}--\ref{fig:voltWtime3}), we use $\epsilon=1\times 10^{-4}$.  

To numerically determine the time point at which the 
transients have decayed in the large-scale simulations, 
we check that the time course of the voltage in a period $T_c$ (inverse of frequency calculated in 
Fig. \ref{fig:cfFreqs} (stars)) is within a specified tolerance of the next cycle; the 
transient time is over only when the voltages match 
\underline{for all} 271 cells:
$$ \text{arg}\min_t \left\{ \frac{\vert v_j(t)-v_j(t+T_c) \vert}{M_k} < \mathbf{e} \left\vert \max_t(v_j(t))-\min_t(v_j(t)) \right\vert, \hspace{.2in}j=1,\dots,271 \text{ cells} \right\}$$
where $M_k=$number of time points in a period, and $\mathbf{e}=1 \times 10^{-3}$ for the two homogeneous tissue and 
$\mathbf{e}=1.7\times 10^{-3}$ for all heterogeneous tissue\footnote{This tolerance 
was the minimum that gave reasonable results when incrementing by $1\times 10^{-4}$ from $\mathbf{e}=1 \times 10^{-3}$. 
One data point corresponding to strong central coupling in the heterogeneous tissue and 
small $\frac{0.5}{15}\text{\,nS}\leq g_{gap} \leq 0.5\,$nS 
(Fig. \ref{fig:trTime}{\bf C} black star) did not yield reasonable transient times for 
similar valued tolerances; see vidx\_Het\_g2\_s2.mp4 in S2.zip.} 
(black stars).  See Figure \ref{fig:trTime}{\bf A, C, E} for comparisons 
of the transient times.

We remark that numerically calculating a precise time when the 
large-scale model reaches a steady-state traveling wave solution 
is tenuous, 
because the results are sensitive to the tolerance.  Yet despite this, we find great qualitative matches 
between the reduced phase model and the large-scale tissue with relatively generic tolerances.  
Indeed, Figure \ref{fig:trTime}{\bf A, C, E} shows 
the match between phase model (thick lines) and large-scale tissues 
(stars connected by lines) can be accurate.  When the 
match is not precise, the following trends are captured: homogeneous peripheral cells (green) have longer transients, followed by homogeneous 
center (blue), then heterogeneous (black); transient times decrease with coupling strength; homogeneous center and heterogeneous tissues 
have similar transient time with stronger coupling.  
The 3 red ovals in Fig. \ref{fig:trTime} left column are for conditions for which boundary effects (i.e., corners of hexagonal grid) 
influence transient times
(see Supplementary Material 
referenced in Fig. \ref{fig:trTime} caption).  
Because our grid is truly 2-dimensional (as opposed to 
radially symmetric grids where nearest neighbor coupling spans disparate Euclidean distances), such boundary effects 
cannot easily be removed.

A common way to quantify the duration of transient dynamics is to consider the \textit{characteristic time-scale}, which for a linear system 
is defined as the inverse of the modulus of the eigenvalue 
with largest (negative) real part 
$\vert \lambda_1 \vert ^{-1}$ of $A$.  This transient time measure does not depend on a particular initial condition or 
numerical tolerance.  
The characteristic time-scale from the reduced phase model is shown in the 
right column of Fig. \ref{fig:trTime}{\bf B, D, F}, 
and it does indeed capture 
all of the trends in both the reduced phase model and large-scale tissues: homogeneous 
peripheral cell tissue has the longest transients, transient time decreases as coupling strength increases, 
heterogeneous tissue has the shortest transient time with weaker coupling, 
homogeneous central cell tissue has similar transient times as the heterogeneous tissues with stronger coupling.  

The results in Figure \ref{fig:trTime} are further detailed with plots of the voltage trajectories, 
along with the transient times calculated both via the reduced phase model 
(cyan arrows, Figs. \ref{fig:voltWtime1}--\ref{fig:voltWtime3}) and 
large-scale simulation (blue arrows, Figs. \ref{fig:voltWtime1}--\ref{fig:voltWtime3}).  
In Figures \ref{fig:voltWtime1}--\ref{fig:voltWtime3}, it is apparent that both transient time estimations from the 
the reduced phase model and the large-scale simulations are relatively accurate.  The only exceptions are perhaps when 
the large-scale simulations exhibit behavior that cannot be well-approximated by a 1D chain of oscillators (i.e., 
plots that correspond to the 3 red ovals in Fig. \ref{fig:trTime} left column).  The constant 
gap junction coupling scheme is shown in Figure \ref{fig:voltWtime1}, with each row consisting of 
a particular cell distribution (i.e., heterogeneous or homogeneous) and each column has the same set of $g_{gap}$ 
values.  We again see that for a given coupling scheme (i.e., column) 
the homogeneous peripheral cell tissue has the longest transient times, 
generally followed by the homogeneous center cell tissue, and then the heterogeneous tissue.  
Figure \ref{fig:voltWtime2} shows the gradient $g_{gap}$ with stronger peripheral coupling, 
and Figure \ref{fig:voltWtime3} is with stronger central coupling.  See Figures S3--S5 (S1Text.pdf) 
for other parameter values not shown in Figures \ref{fig:voltWtime1}--\ref{fig:voltWtime3}.

Thus far, we have compared the large-scale models with its corresponding 
reduced phase model.  We now use the phase model to investigate the 
spatio-temporal dynamics for the heterogeneous 
networks in a large region of parameter space.  
For all 3 coupling schemes, we vary both the initial condition and 
gap junction coupling values.  We consider 
21 initial conditions varying gradually, 
i.e., where the system starts primed for center-to-peripheral 
traveling waves (Fig.\ref{fig:lngPhase}, i in gray inset), to 
starting at complete synchrony (Fig.\ref{fig:lngPhase}, iii), to 
being primed for peripheral-to-center traveling waves (Fig.\ref{fig:lngPhase} 
v).  We choose 391 gap junction coupling values (or sets) that span a 
large range \citep{verheule01}.  
The black stars correspond to the large-scale Severi model simulations that have been investigated.  
Note that simulation of the large-scale pacemaker tissue model over this wide parameter regime for gap junctional values and initial conditions would require a prohibitively large amount 
of computational resources.

In this study of the reduced model, 
we first determined if the system settled to 
a periodic solution (i.e., traveling wave that satisfies 
Eqs.\eqref{eqn:redModel}--\eqref{eqn:exist_tw}).  
With constant gap junction coupling (Fig.\ref{fig:lngPhase}{\bf A}) 
and with stronger central 
gap junction (Fig.\ref{fig:lngPhase}{\bf C}), traveling wave solutions exist 
and are peripheral-to-center (green) 
for all $g_{gap}$ and initial conditions considered.  
With stronger peripheral gap junction (Fig.\ref{fig:lngPhase}{\bf B}), 
peripheral-to-center traveling waves 
exist for almost all parameters, except for very small 
$g_{gap}$ values.  The reduced model predicts that for very weak 
coupling, periodic solutions do not exist (blue), which 
we investigated further below (see Fig.\ref{fig:nonTW}{\bf A,B,C}).  
Interestingly, there is a region where the peripheral-to-center 
traveling waves exist but have such large frequencies that the 
subsequent waves initiate before the previous one finishes 
(yellow in Fig.\ref{fig:lngPhase}{\bf B}), 
resulting in phase lags that are non-monotonic; we  
investigated this response further in Fig.\ref{fig:nonTW}{\bf D,E,F}.  
The tissue frequency when periodic solutions exist 
(Fig.\ref{fig:lngPhase}{\bf D,E,F}, same color scale) show no dependence 
on initial condition and decrease with coupling strength 
(consistent with Fig.\ref{fig:cfFreqs}).  These results 
validate our assertion that 
the traveling waves are locally stable (Fig.\ref{fig:stability},S1) for 
many initial conditions.  
The right column (Fig.\ref{fig:lngPhase}{\bf G,H,I}, same color scale) 
shows how the 
transient times depend on initial condition and coupling strength 
(lower range of $g_{gap}$ shown because times are minuscule with larger 
$g_{gap}$).  Transient times decrease with coupling 
strength and also decrease as the initial conditions are `closer' 
to peripheral waves (bottom of y-axis), which is the stable periodic 
solution (except for blue and yellow in {\bf B}).  
Thus, we have used the reduced phase model to both verify key aspects of 
the spatio-temporal dynamics over a large region of parameter space and identify regimes of irregular behavior, investigated further below.  

\begin{figure}
\centering
\includegraphics[width=\textwidth]{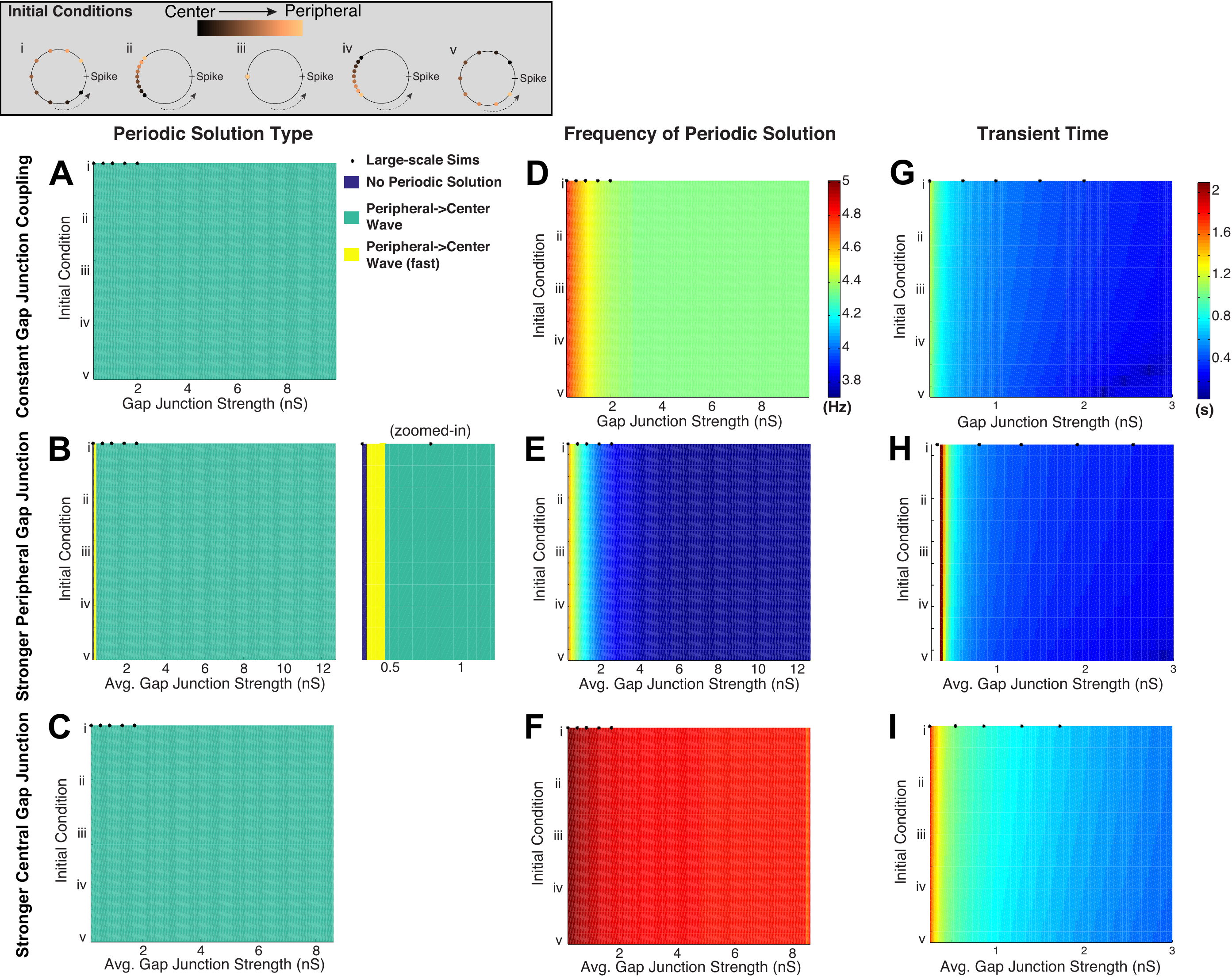}
\caption{\label{fig:lngPhase} Phase model results with 
many initial conditions and coupling strengths for 
heterogeneous tissues.  Top panel (gray): illustrative subset of the 21 initial conditions 
we consider, gradually varying from being 
primed for center-to-peripheral traveling 
waves (i), to starting at synchrony (iii), to being primed for 
peripheral-to-center waves (v).  The large-scale simulations 
we previously considered correspond to the five 
black stars ({\bf A--I}).  With constant gap 
junction coupling ({\bf A}) and 
with stronger central 
gap junction ({\bf C}), pertipheral-to-center 
traveling wave solutions exist (green) 
for all $g_{gap}$ sets and initial conditions considered.  
{\bf B}) With stronger peripheral coupling, peripheral-to-center 
traveling waves exist for almost all parameters, the exception 
is with smaller coupling values where periodic solutions either 
do not exist (blue) or have very fast frequencies with 
non-monotnic phase lags (yellow, see Fig.\ref{fig:nonTW}{\bf D,E,F} 
for further details).  
The tissue frequency when periodic solutions exist 
for all 3 coupling schemes ({\bf D,E,F}), same color scale) 
show no dependence 
on initial condition and decrease as coupling strength increases 
(cf. Fig.\ref{fig:cfFreqs}).  The transient times ({\bf G,H,I}, 
same color scale) decrease as coupling strength increases, and also 
decrease as initial conditions vary from (i) to (v) because 
the system starts 'closer' to peripheral-to-center traveling waves 
(i.e., the stable periodic solution).
}
\end{figure}

\begin{figure}
\centering
\includegraphics[width=\textwidth]{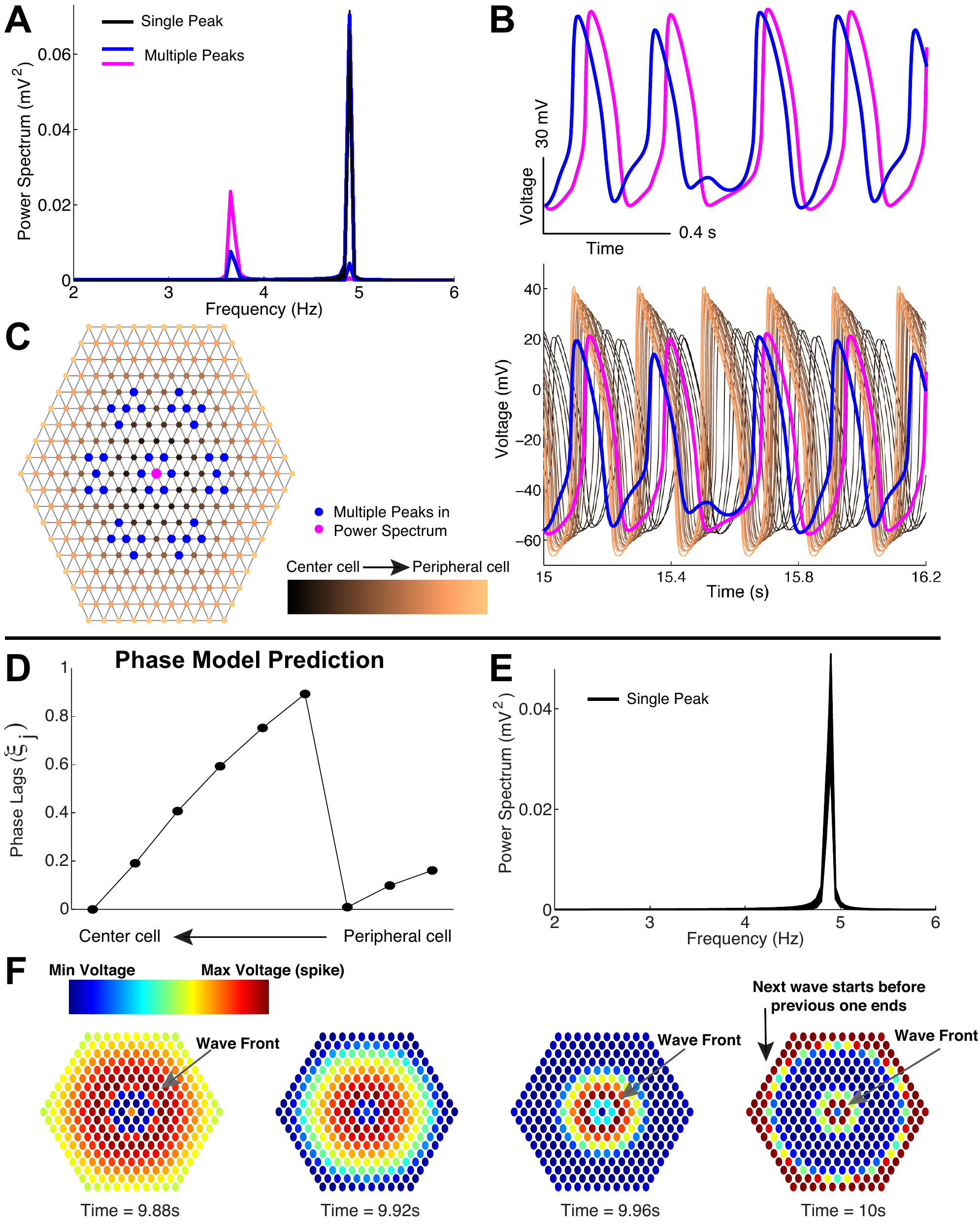}
\caption{\label{fig:nonTW} Irregular behavior in phase 
models correspond to irregular behavior in the large-scale models.  
A tissue structure and coupling scheme, for which a traveling wave solution does not exist {\bf A--C}.  
The heterogeneous tissue has stronger peripheral gap junction coupling, with $\frac{0.5}{15}\text{\,nS}\leq g_{gap} \leq 0.5\,$nS.
{\bf A}) Power spectrums of all 271 voltage traces (after transients have decayed, from 10\,s--20\,s).  
There are cells (37 total in blue and magenta) with multiple peaks in the power spectrum, a signature of 
multiple time-scales; in contrast to other parameters considered.  
{\bf B}) Two types of voltage traces that have irregular dynamics (different spike widths, heights, etc. from cycle to cycle); 
all but the center most cell (magenta) have the blue voltage traces.  
{\bf C}) The spatial location of the cells with multiple time-scales in voltage.  See Supplementary Material (vidx\_Het\_g2\_s1.mp4 in S2.zip).  
{\bf D)--F)} Again with strong peripheral gap junction coupling but 
with $\frac{0.6}{15}\text{\,nS}\leq g_{gap} \leq 0.6\,$nS, corresponding 
to the yellow region in Fig.\ref{fig:lngPhase}{\bf B}.  The phase 
model predicts peripheral-to-center waves that are so fast that the 
phase lags are non-monotonic ({\bf D}).  {\bf E}) The power spectrum of 
all cells in the corresponding large-scale simulation has 1 peak, 
like all other tissues with traveling waves.  
{\bf F}) Snapshots of the large-scale model in time show 
that the phase model prediction is verified, show peripheral-to-center 
traveling waves are so fast that the next wave starts before the prior 
one ends.  See Supplementary Material (vidx\_Het\_gSMp6\_s1.mp4 in S2.zip).
}
\end{figure}

From the large-scale tissue simulations, we identified a single parameter set for which a traveling wave solution likely did not exist, 
a heterogeneous tissue with stronger peripheral gap junction coupling. The reduced model further predicted a regime of parameter conditions for which traveling wave solutions 
likely do not exist and for which the single parameter set falls within, $\frac{0.5}{15}\text{\,nS}\leq g_{gap} \leq 0.5\,$nS 
(see Fig. \ref{fig:lngPhase}{\bf B} in blue).  The large-scale tissue simulation  with
heterogeneous tissue and stronger peripheral gap junction coupling thus validates the reduced model prediction of irregular dynamics.  In this tissue, we find that the 
power spectrum of all 271 voltage traces (after transients have decayed) have a subset of 37 cells 
with multiple peaks (Fig. \ref{fig:nonTW}{\bf A}).  
This suggests that the voltage time course for some cells have multiple time-scales, in contrast to previously described tissues thus far, for which the voltage trace power spectrum is unimodal for all cells.  Inspection 
of the voltage traces indeed shows irregularities from cycle to cycle (Fig. \ref{fig:nonTW}{\bf B}), with 
different spike widths, heights, etc. for the subset of cells with multi-modal power spectrums.  The voltage 
trajectories of these `irregular' cells are all synchronized except for the center most cell 
(see Fig. \ref{fig:nonTW}{\bf C} for the spatial location).  Importantly, the failure to find a traveling wave 
solution in the reduced model is consistent with the large-scale simulations, where the persistent solution 
is indeed not a traveling wave solution,  demonstrating that the reduced model formulation can provide an approach to identify regimes of irregular voltage dynamics.  

Recall the yellow region in Figure \ref{fig:lngPhase}{\bf B}
(strong peripheral coupling) where the 
heterogeneous phase model predicts a very fast peripheral-to-center 
traveling wave, such that the phase lags are non-monotonic 
(Fig.\ref{fig:nonTW}{\bf D}).  We verified this prediction with 
corresponding large-scale simulations for 
$\frac{0.6}{15}\text{\,nS}\leq g_{gap} \leq 0.6\,$nS.  
The power spectrum of the 
voltage (after transients have decayed) for all 
271 cells have a single peak (Fig.\ref{fig:nonTW}{\bf E}), verifying 
the existence of a periodic solution.  Finally, Figure \ref{fig:nonTW}{\bf F} 
shows snapshots of the voltage (scaled by max and minimum voltage 
to compare across different cells) that demonstrate peripheral-to-center traveling waves, with dynamics such that the next wave initiates before the previous one ends, as predicted by the reduced phase model.

\subsection{Effects of Varying Gap Junction Coupling}

Above we described how the precise distribution of gap junction coupling 
strengths is still unknown despite ultrastructure studies because of the 
irregular structure and other components in SAN, which motivated us to 
consider the three different coupling schemes.  Thus far we have chosen 
modest gap junction values (up to 4\,nS for some pairs) in the large-scale 
simulations that have enabled analysis of how coupling values alter existence 
of traveling waves, tissue frequency, transient times, and have enabled 
comparisons of these entities across tissue models with different cell 
types.  Experiments by \citet{verheule01} in rabbit SAN report an average 
$g_{gap}$ value of 7.5\,nS, with the median value among their 32 pairs of 
5.3\,nS and a range from 0.6 to 25\,nS.  However, these measurements are 
performed in intact cell pairs and does not account for effectively reduced 
coupling between cells in the setting of highly fibrous SAN tissue. 
Reduced coupling can be further exacerbated with age and SAN 
disease \citep{jones2004ageing}. Therefore, we perform additional 
large-scale simulations over a wide range of coupling strengths; for 
constant gap junction coupling we set $g_{gap}$=5, 7.5, 10\,nS, 
for both types of gradient coupling 
(stronger central or stronger peripheral) the values varied from 
$\frac{10}{15}\,\text{nS} \leq g_{gap} \leq 10\,\text{nS}; 
1\,\text{nS} \leq g_{gap}\leq 15\,\text{nS}; \frac{20}{15}\,\text{nS} 
\leq g_{gap} \leq 20\,\text{nS}$. 

Simulations are compared with the corresponding reduced phase model  (see 
Fig. S6) as in Figures \ref{fig:cfFreqs} and \ref{fig:trTime} 
even though the approximations rely on weak coupling, 
and our prior results still qualitatively hold (i.e., reduced model 
qualitatively and sometimes quantitatively captures the trend of tissue 
frequency).  The tissue frequencies for the heterogeneous tissue saturate 
with larger coupling values (Fig. S6A--C), while the transient times are very 
small for all tissue types (Fig. S6D--F).  The detailed voltage trajectories 
for these larger $g_{gap}$ values are shown in Figs. S7--S9.  Other large coupling 
values are shown here for completeness despite violating assumptions in the 
reduced phase model framework, but our main focus is on small to moderate 
coupling values that show more nuanced dynamics and entities (voltage, tissue 
frequency) consistent with experiments.

\section{Discussion}

In this study, we demonstrate the utility of the mathematical framework of coupled phase oscillators 
to mechanistically characterize the spatio-temporal dynamics 
of a large locally coupled tissue of heterogeneous 
sinoatrial node (SAN) cells, for which 
the individual cell models themselves are complex with many state variables and compartments.  Summarizing our approach and main findings, we implemented a recently developed model of the central 
pacemaker cells in the SAN \citep{severi12} and modified model parameters to capture a heterogeneous population of
pacemaker cells to yield a large-scale SAN tissue tissue model.  
In electrically (gap junction) 
coupled tissue consisting of 
hundreds of pacemaker cells, we observed vastly different `steady-state' 
or persistent behavior for tissue comprised of different  intrinsic cell types.  
In these models, we found that the persistent 
behavior (synchrony for homogeneous 
cell tissue or peripheral-to-center waves for heterogeneous tissue) 
was highly robust, even as both coupling schemes and 
gap junction strengths varied.  

To explain observations in the large-scale models, 
we applied a reduced model framework consisting of 
a 1D chain of phase oscillators that accounted for different 
cell types and coupling strengths from the full 2D tissue.  
Importantly, the reduced phase model relied on quantities 
(i.e., intrinsic frequencies, cell-to-cell interactions) 
from the individual pacemaker cell model but did not rely on 
computationally expensive simulations of the large-scale tissue model.  

The key novel results and insights of our study can be summarized as follows:
1) The reduced phase model framework successfully captured when traveling waves 
exist, uncovered the stability of these 
persistent states, and was used to calculate the 
transient times -- all in a variety of tissue 
types, and coupling schemes and strengths.  2) In addition to requiring less computational resources to capture the large-scale tissue phenomena, the phase models facilitate detailed mathematical analysis and wider parameter studies of initial conditions and coupling schemes. 3) Despite incorporating a highly detailed biophysical model of SAN dynamics at the individual cell level, the large-scale tissue simulations predict that center-to-peripheral wave generation is only a transient behavior. Additionally, a wide parameter study in the reduced phase model further predicts that this finding holds, with the duration of transients depending on coupling strength and initial conditions.  This demonstrates that additional model components are necessary to reproduce the physiologically observed center-to-peripheral wave generation. 
Note that experiments on rabbit SAN where the atrium is removed show 
peripheral-to-center traveling waves \citep{kirchhof87}, consistent with 
our heterogeneous model (also see \citet{kodama85,opthof87intrinsic} 
for experiments that further validate our peripheral traveling wave results).  
Our future work will focus on addressing this model limitation, considering two potential mechanisms: i) incorporating additional anatomical details, specifically the electrical coupling between the SAN and the atria, and ii) incorporating spatially variable sensitivity to autonomic regulation that regulates pacemaker intrinsic frequency.

Phase reduction methods to explore dynamics of SAN models has a long history, 
albeit with less physiological models than the \citet{severi12} 
models we consider here.  Many of the works we review here exploit 
simple models using a \textit{phase-resetting/response curve} (PRC) 
or equivalent entity that quantifies how weak perturbations alter the time to 
the next action potential.  
The phase models we employ in this paper depend on the PRC, $\Delta$, 
but the method of averaging is used to yield interaction functions 
$H$ (see $\Delta$ in Eq.\eqref{eqn:Hfunc}).   \citet{ikeda82} 
and colleagues used minimal models with PRC functions based on experimental 
recordings of myocardial pacemaker cells in mammals (rabbit, cat) to 
mathematically analyze how pairs of cells can result in the existence of 
synchronized solutions -- with simulations, they also considered 1D chains 
and 2D `sheets' of pacemaker cells.  \citet{ikeda86} and 
\citet{deBruin83} used simulations to probe how altering PRCs shape rhythms 
and synchrony.  Based on simulations, Abramovich-Sivan and Akselrod 
studied the spatiotemporal patterns of a minimal 
SAN model \citep{abramovich2000} with a PRC similar in form to Ikeda and 
colleagues; they also studied how the location and duration of traveling 
waves change with different intrinsic cycle lengths and PRCs 
\citep{abramovich99a}, and how the atrium alters activity patterns 
\citep{abramovich99b}.  \citet{ostborn01} employed a circle map 
simplification (i.e., no differential equations) of simple SAN models to 
determine entrainment dynamics and different locking states (i.e., n:m 
locking, beyond traveling waves).  {\"O}stborn and colleagues studied how 
coupling strengths can lead to entrainment of cells with different cycle 
lengths \citep{ostborn02,ostborn04}, rather than how various PRC shapes alter 
entrainment.  

A number of researchers have considered models with different intrinsic cycle lengths \citep{abramovich99a,abramovich99b,abramovich2000,ikeda82,ikeda86,deBruin83,ostborn01,ostborn02,ostborn04,mirollo90,acebron05,mazurov2006rhythmogenesis}.  
These studies have typically used a `pulse-coupling' approximation (effects 
of coupling only occur at one time point of the action potential) to model 
gap junction coupling, whereas our analysis does not require this.  The main 
advantage of `pulse-coupling' is apparently to have a well-defined set of 
coupled maps that (perhaps) are easier to analyze, rather than having a 
system of differential equations where the effects of coupling persists even 
when cells are not spiking.  Another difference is that these studies
\citep{abramovich99a,abramovich99b,abramovich2000,ostborn01,ostborn02,ostborn04,ikeda82,ikeda86,deBruin83} assume the PRCs can be manipulated in any 
desired way, whereas PRCs (and interaction functions) computed from biophysically detailed models have a complex dependence on parameters -- they 
cannot be precisely controlled.  Note that there may be other topological 
properties of the PRC when accounting for cell type that could be explored in 
the future \citep{guevara90,ikeda86,deBruin83}.  
In contrast to these referenced works, we have focused on quantitative 
accuracy of the phase-reduced models with the large-scale models, as well as 
the shortcomings of the reduced phase models.

While the general analysis we used here has 
been applied to minimal cardiac cell models 
(i.e., see section 14.2 of \cite{keenerSneyd}), 
to our knowledge, our analysis of a complex and highly detailed biophysical 
representation of pacemaker cells is unique and novel.  Thus, our 
application is valuable because it establishes a framework for future 
mechanistic investigations of irregular SAN electrical function, driven by 
defective ion channel gating or expression. Ion channel dysfunction (e.g., 
reduced funny current associated with sick sinus syndrome 
\citep{schweizer2014symptom}) can be incorporated into the underlying 
pacemaker cell models and subsequent reduced phase model stability can be 
predicted. As discussed in the previous paragraph, the application of 
reduced phase models with direct comparisons to 
large-scale heterogeneous tissue 
has been scarce; see 
\citep{Ly_heterOsc_14,zhou13,ashwin16} for some exceptions.  Regarding 
heterogeneity with the SAN tissue model, in terms of both cell type and cell-cell coupling gradients, we note that the large-scale model of 271 cells 
represents a subset of the full SAN tissue, and therefore a limitation of our 
model is that the steepness of parameter gradients is larger than that of a 
full-sized tissue. However, we chose to consider the full range of parameter 
conditions present in the full-size tissue (i.e., center-to-peripheral cell 
gradients, and cell-cell coupling gradients), a trade-off at the expense of a 
steeper gradient.  


Our work could potentially elucidate large simulation studies in which center-to-peripheral traveling waves are observed.  
    We do observe central-to-peripheral traveling waves, but only transiently, and the transient times strongly 
    depend on cell types  and coupling within the tissue.  Specifically, 
    the transients times were generally longer for homogeneous tissue than heterogeneous.  
    In principle, this framework could be applied to other models 
    \citep{zhang2000,oren2010} to mathematically characterize the steady-states and transient times.  
    We note that \citet{oren2010} argued that more realistic center-to-peripheral traveling waves only occurred in homogeneous 
    tissue, and not in heterogeneous tissue 
    (\citet{joyner86} also showed this in a homogeneous network).  
    Other modelers have shown 
    that center-to-peripheral waves with heterogeneous cells when SAN 
    is coupled to the atrium: \citet{inada14} showed this with the Kurata 
    model, see \citet{unudurthi14} for a review.

We have seen that there are regimes for which traveling wave solutions do not 
exist in the phase model (nor in the large-scale model) and 
the cell behavior is irregular from cycle to cycle (Fig. \ref{fig:nonTW}).  Extending the phase oscillator framework to these complex dynamics 
and beyond is a promising area of future research, especially given 
the implications for these dynamics on cardiac arrhythmias and other 
disease. We are also interested in developing reduced phase model methods 
to analyze cardiac SAN tissue models that incorporate additional physiological features: further SAN cell 
population heterogeneity, including spatial heterogeneity in sensitivity to 
autonomic regulation and shifts in pacemaker origin \citep{csepe16}; incorporating  both regions of electrical isolation and preferential conduction pathways between the SAN and the atria that have been identified in human and canine SAN  \citep{fedorov12}; and including coupling with fibroblasts. Consistent with prior computational studies 
\citep{garny03}, we expect that the hyperpolarizing 
influence of the atrium would reduce the frequency of the intrinsically 
faster peripheral cells. Developing the mathematical analysis for a phase 
model with coupling between oscillatory pacemaker cells, excitable atrial 
cells, and  non-excitable 
fibrous tissue is also a novel area for future work. 
Although phase oscillator theory is well-developed 
\citep{ashwin16}, and there is some theoretical 
progress in defining a phase of an excitable cells 
\citep{thomas14,schwabedal13}, a unified theory to succinctly 
describe oscillators coupled to excitable tissue is lacking.  
We hypothesize that these additional model details may yield stable center-
to-peripheral traveling waves and also shifting origins of pacemaker rhythms 
in response to autonomic activity. 

As experimental techniques advance and single cell models naturally 
increase in complexity 
with more variables and parameters, the necessity of tractable 
and insightful models will be dire.  
We have demonstrated that these 
reduced phase models and corresponding mathematical analysis are 
valuable for studying large realistic coupled cardiac tissues.  
For example, as more accurate data 
become available about the distribution of pacemaker cells' intrinsic frequency and cell coupling, this 
framework can be developed and extended 
to probe the spatio-temporal dynamics efficiently rather than 
performing computationally 
intensive simulations for a wide range of parameters.

\appendix

\section{Detailed equations for the large-scale model}\label{sec:pm_detail}
 
Here we provide more details for the full pacemaker cell models (Eq. (\ref{eqn:volt})).  For the complete details, we provide the Matlab code for the models, publicly available at 

\hspace{-.24in} 
{\tt http://github.com/chengly70/SanHeteroSeveri}.  The complete 
details of the center cell pacemaker model is provided in 
\citet{severi12}, with computer code publicly available at 

\hspace{-.24in}
{\tt http://models.cellml.org/e/139/severi\_fantini\_charawi\_difrancesco\_2012.cellml/view}.  
There are 
31 total dynamic variables with their own differential equation for \textit{each} cell, and many auxiliary functions.

The voltages are in units of milli-Volts (mV).  The sodium current is:
\begin{eqnarray}
 	I_{Na}(v) &=& g_{Na} m^3 h (v-E_{Na}) \\
	E_{Na} 	  &=& \frac{RT}{F} \ln\left( Na_o/ Na_i \right) \\
	\frac{d x}{dt} &=& \alpha_x(v)(1-x) - \beta_x(v)x, \hbox{    where }x\in\{m, h\}
 \end{eqnarray}
where $g_{Na}$ is the conductance (varies, see Table \ref{tab:hetparm}), 
$m$ and $h$ are the sodium activation and inactivation variables, $R$ is the universal gas constant, $F$ is Faraday's constant, $T$ is temperature in Kelvin (310, or 36.85$^\circ$ C), 
$Na_o=140$ mM is the extracellular sodium concentration; $Na_i$ is the dynamic intracellular concentration with its own differential equation; 
$\alpha_m(v)$, $\alpha_h(v)$, $\beta_m(v)$, $\beta_h(v)$ are standard gating variables.

The funny current is:
\begin{eqnarray}
 	I_f(v) &=& g_{f,Na} y_f^2 K_o/(K_o+45)(v-E_{Na}) + g_{f,K} y_f^2 K_o/(K_o+45)(v-E_{K}) \\
	E_{K} 	  &=& \frac{RT}{F} \ln\left( K_o/ K_i \right) \approx -86.96\,\hbox{mV} \\
	\frac{d (y_f)}{dt} &=& (y_{f,\infty}(v)-y_f)/\tau_{yf}(v)
 \end{eqnarray}
where $g_{f,Na}$, $g_{f,K}$ are the conductances (varies, see Table \ref{tab:hetparm}); 
$K_o=5.4$ mM is the extracellular potassium concentration, $K_i=140$ mM is the intracellular potassium concentration.  

The transient outward potassium current is:
\begin{eqnarray}
 	I_{to}(v) &=& g_{to} q*r*(v-E_{K}) \\
	\frac{d x}{dt} &=&  (x_\infty(v)-x)/\tau_{x}(v), \hbox{    where }x\in\{q, r\} 
\end{eqnarray}
with $g_{to}=0.002\,\mu$S; for this and other currents, the steady state functions $x_\infty(v)$ and time constant $\tau_{x}(v)$ are standard 
(see \citet{severi12}).  

The L-type calcium current is:
\begin{eqnarray}
 	I_{Ca,L} &=& I_{si,Ca}+I_{si,K}+I_{si,Na} \\
	I_{si,Ca} &=& \frac{2 P_{CaL}* v}{\frac{RT}{F} \left(1-\exp\left(-2v\frac{F}{RT}\right) \right)}\Big( Ca_{sub}-Ca_o\exp\left(-2v\frac{F}{RT}\right) \Big)*d_L*f_L*f_{Ca} \\
	I_{si,K} &=& \frac{3.65\times10^{-4} P_{CaL}* v}{\frac{RT}{F} \left(1-\exp\left(-v\frac{F}{RT}\right) \right)}\Big( K_i-  K_o\exp\left(-v\frac{F}{RT}\right) \Big)*d_L*f_L*f_{Ca} \\
	I_{si,Na} &=& \frac{1.85\times10^{-5} P_{CaL}* v}{\frac{RT}{F} \left(1-\exp\left(-v\frac{F}{RT}\right) \right)}\Big( Na_i-  Na_o\exp\left(-v\frac{F}{RT}\right) \Big)*d_L*f_L*f_{Ca} \\
	\frac{d x}{dt} &=&  (x_\infty(v)-x)/\tau_{x}(v), \hbox{    where }x\in\{d_L, f_L\} \\
	\frac{d (f_{Ca})}{dt} &=&  (f_{Ca,\infty}(Ca_{sub})-f_{Ca})/\tau_{fCa}(Ca_{sub})
\end{eqnarray}
where $P_{CaL}$ is the conductance (varies, see Table \ref{tab:hetparm}), $Ca_o=1.8$ mM is the extracellular calcium concentration, $Ca_{sub}$ is the dynamic subspace calcium concentration 
with its own differential equation.  

The T-type calcium current is:
\begin{eqnarray}
		I_{Ca,T} &=& \frac{2 P_{CaT}* v}{\frac{RT}{F} \left(1-\exp\left(-2v\frac{F}{RT}\right) \right)}\Big( Ca_{sub}-Ca_o\exp\left(-2v\frac{F}{RT}\right) \Big)*d_T*f_T \\
	\frac{d x}{dt} &=&  (x_\infty(v)-x)/\tau_{x}(v), \hbox{    where }x\in\{d_T, f_T\} 				 
\end{eqnarray}
where $P_{CaT}$ is the conductance (varies, see Table \ref{tab:hetparm}).

The rapid delayed rectifier K$^+$ and slow delayed rectifier K$^+$ currents are:
\begin{eqnarray}
		I_{Kr} &=& g_{Kr}* (0.9*p_{aF} + 0.1*p_{aS})*p_i* (v-E_{K})\\
	I_{Ks} &=& g_{Ks}* n^2* (v-E_{K})\\
	\frac{d x}{dt} &=&  (x_\infty(v)-x)/\tau_{x}(v), \hbox{    where }x\in\{p_{aF}, p_{aS}, p_i, n\} 				 
\end{eqnarray}
where both $g_{Kr}$ and $g_{Ks}$ conductances vary by cell type (see Table \ref{tab:hetparm}).


The sodium-potassium pump current $I_{Na,K}$ and sodium-calcium exchanger current $I_{Na,Ca}$ are:
\begin{eqnarray}
		I_{Na,K} &=& \frac{g_{NaK}}{(1+1.4/K_o)^{1.2}(1+14/Na_i)^{1.3}(1+\exp(-(v-E_{Na}+110)/20))} \\
		I_{Na,Ca} &=& g_{NaCa}\frac{x_2 k_{21}-x_1 k_{12}}{x_1+x_2+x_3+x_4}
\end{eqnarray}
The components $x_1,\dots,x_4$ and $k_{21}$, $k_{12}$ represent the detailed calcium dynamics in the sarcoplasmic recticulum (SR), myoplasm, junctional SR, subspace SR, as well as outside 
of the cell.  We omit the rest of the equations and functions due to their complexity and point the 
interested reader to our freely available computer code.  We alter cell length and cell radius for different cell types, that in turns effects the cell volume, submembrane space volume, myoplasmic volume, junctional SR volume, network SR volume, which all effect the calcium dynamics.

The values for $g_{NaK}$ and $g_{NaCa}$ for the 9 different cells were manually altered to insure ion homeostasis and a stable limit cycle in the uncoupled cells, the values are listed in Table \ref{tab:pumpVals}.


\section*{Acknowledgements}
The authors acknowledge and thank the Virginia Commonwealth University Center for High Performance Computing (CHiPC)
for use of the Teal cluster to perform numerical simulations.  CL is supported by a grant from the Simons Foundation (\# 355173).  

\section*{Supplementary Material}

\textit{S1Text.pdf --  Supplementary Figures }

A PDF files that contains 9 extra figures that expound on 
Figures \ref{fig:stability}, \ref{fig:trTime}{\bf B,D,F}, 
\ref{fig:voltWtime1}--\ref{fig:voltWtime3}, and show the 
results for large gap junction strengths.  

\vspace{.1in}

\textit{S2.zip --  Movies }

A zip file containing several movie files (.mp4) showing the 
voltage evolution on the 2D hexagonal grid.  Naming convention: 

\texttt{vidx\_[Het/HomC/HomP]\_io\_g[2/22/3/33/4]\_s[1/2/3].mp4} 

\hspace{-.24in} 
\texttt{Het} is with all cell types (heterogeneous), \texttt{HomC} 
is a homogeneous tissue with all center cells, \texttt{HomP} 
is a homogeneous tissue with all peripheral cells. 

\hspace{-.24in}
\texttt{s1}: gradient gap strength with peripheral having 15X larger strength than center. 

\hspace{-.24in}
\texttt{s2}: gradient gap strength with center having 15X larger strength than peripheral. 

\hspace{-.24in}
\texttt{s3}: constant gap strength throughout.

\hspace{-.24in}
\texttt{g2}=0.25\,nS ([0.5/15,0.5]\,nS range of $g_{gap}$ when gradient gap 
strength s1 \& s2)

\hspace{-.24in}
\texttt{g22}=0.625\,nS ([1.25/15,1.25]\,nS range with gradient gap schemes)

\hspace{-.24in}
\texttt{g3}=1\,nS ([2/15,2]\,nS range with gradient gap schemes)

\hspace{-.24in}
\texttt{g33}=1.5\,nS ([3/15,3]\,nS range with gradient gap schemes)

\hspace{-.24in}
\texttt{g4}=2\,nS ([4/15,4]\,nS range with gradient gap schemes)

Also two movie files,  
\texttt{vidx\_Het\_gSMp[6,7]\_s1.mp4}, correspond to 2 points 
in the yellow region of Figure \ref{fig:lngPhase}{\bf B} with initial 
condition (i).

\section*{References}
\bibliographystyle{elsarticle-harv} 

\end{document}